\documentclass[draftcls, onecolumn]{IEEEtran}

\usepackage{graphicx}
\usepackage{amsmath}
\usepackage{latexsym}
\usepackage{amssymb}
\usepackage{bm}
\usepackage{amsmath}
\usepackage[caption=false,font=footnotesize]{subfig}

\newcounter{dclt}

\newcommand{\argmax}{{\mathrm{argmax}}}

\newcommand{\Hb}{{\mathbf{H}}}
\newcommand{\vb}{{\mathbf{v}}}
\newcommand{\hb}{{\mathbf{h}}}

\newcommand{\ub}{{\mathbf{u}}}

\newcommand{\nn}{\nonumber}
\newcommand{\eqqq}{\!\!\!\!&=&\!\!\!\!}
\newcommand{\leqqq}{\!\!\!\!&\leq&\!\!\!\!}

\newtheorem{theorem}{Theorem}

\newtheorem{lemma}{Lemma}

\newtheorem{proposition}{Proposition}
\newtheorem{remark}{Remark}

\begin{document}

\title{Analysis of Feedback Overhead for MIMO Beamforming over Time-Varying Channels}

\author{\large Jaewon Kim and Jonghyun Park}

\maketitle

\begin{abstract}
In this paper, the required amount of feedback overhead for multiple-input multiple-output (MIMO) beamforming over time-varying channels is presented in terms of the entropy of the feedback messages. In the case that each transmit antenna has its own power amplifier which has individual power limit, it has been known that only phase steering information is necessary to form the optimal transmit beamforming vector. Since temporal correlation exists for wireless fading channels, one can utilize the previous reported feedback messages as prior information to efficiently encode the current feedback message. Thus, phase tracking information, difference between two phase steering information in adjacent feedback slots, is sufficient as a feedback message. We show that while the entropy of the phase steering information is a constant, the entropy of the phase tracking information is a function of the temporal correlation parameter. For the phase tracking information, upperbounds on the entropy are presented in the {\em Gaussian entropy} and the {\em von-Mises entropy} by using the theory on the maximum entropy distributions. Derived results can quantify the amount of reduction in feedback overhead of the phase tracking information over the phase steering information. For application perspective, the signal-to-noise ratio (SNR) gain of phase tracking beamforming over phase steering beamforming is evaluated by using Monte-Carlo simulation. Also we show that the derived entropies can determine the appropriate duration of the feedback reports with respect to the degree of the channel variation rates.
\end{abstract}

\renewcommand{\thefootnote}{\fnsymbol{footnote}}

\section{Introduction} \label{sec:Intro}

Consider a multiple-input multiple-output (MIMO) time-varying wireless channel for $n$-th frame instance:
\begin{equation}
\mathbf{y}[n]= \Hb[n] \vb[n] x[n] + \mathbf{w}[n] \label{def:y[n]}
\end{equation}
where the scalar $x[n]$ represents the transmit data symbol, the matrix $\Hb[n]$ represents $N_R \times N_T$ wireless channel with
the number of receive antennas $N_R$ and the number of transmit antennas $N_T$, $\vb[n]$ is an $N_T \times 1$ transmit beamforming
vector, the vector $\mathbf{y}[n]$ represents received signals, and $\mathbf{w}[n]$ is the additive independent and identically
distributed (i.i.d.) complex-valued noise vector with zero mean and unit variance on each of its components. Each entry of the channel
matrix $\Hb[n]$ is assumed to be an i.i.d. complex Gaussian random variable (RV) with zero mean and unit variance. The optimal
beamforming scheme in terms of maximizing the output signal-to-noise ratio (SNR) is known as \emph{eigenbeamforming} which utilizes the
dominant eigenvector of the channel matrix as a transmit beamforming vector and the maximum ratio combining vector at the receiver
side~\cite{Goldsmith05}. Since the eigenvector is a unit norm vector (not a unit elemental power on every entry of the vector), the optimal
transmit vector satisfies the total power constraint which limits the sum of transmit powers less than a given constant. Utilization
of the optimal transmit beamforming method, however, can be limited by practical constraints such as limited data rate feedback
links~\cite{Au-Yeung07}--\cite{Love05} and transmit power constraints imposed at the transmitter~\cite{Yu07}.

Problem known as \emph{limited feedback communication} has been an active area of research, and one of main ideas to cope with this problem is to use codebook-based precoding methods~\cite{Love08}. Instead of using the channel state information itself for signal processing at the transmitter side, the codebook-based methods only feedback the indices of favorable beamforming vectors to the transmitter, which are already known to both the transmitter and the receiver. In recent years, there have been intensive studies on codebooks construction, such as generalized Lloyd-type vector quantization methods~\cite{Au-Yeung07},~\cite{Roh06} and Grassmannian line packing methods~\cite{Love03G},~\cite{Love05}.

Instead of considering the total power constraint, imposing per-antenna power constraint (PAPC) is more realistic in practice, since each transmit antenna has its own power amplifier which has individual power limit~\cite{Yu07}. Under the PAPC, it is proved that only phase steering values are necessary (without magnitude information) to form the optimal transmit beamforming vector, regardless of the number of receive antennas~\cite{Murthy07}. This is also referred to as equal gain transmission (EGT), which utilizes identical power but different phase steering values across transmit antennas~\cite{Love03E}. Thus we consider beamforming vector satisfying PAPC
\begin{equation}
\vb[n] = \frac{1}{\sqrt{N_T}}
\left[\begin{IEEEeqnarraybox*}[][c]{,c/c/c/c,} e^{j\theta_0[n]} &
e^{j\theta_1[n]} & \cdots & e^{j\theta_{N_T-1}[n]}
\end{IEEEeqnarraybox*}\right]^T
\label{def:v[n]}
\end{equation}
where $\theta_i[n]$ denotes the phase steering value of the $i$-th transmit antenna \cite{Heath98}--\cite{Xia05}. The optimal phase steering values in terms of maximizing the output SNR are determined by
\begin{equation}
\bm{\theta}[n] = \underset{\theta_i[n]}\argmax \|\Hb[n]\vb[n]\|^2
\label{eq:Theta[n]_argmax}
\end{equation}
where $\bm{\theta}[n]$ is the optimal phase steering vector defined as
\begin{eqnarray}
\qquad\bm{\theta}[n] =
\left[\begin{IEEEeqnarraybox*}[][c]{,c/c/c/c,}
                 0 & \theta_1[n] & \cdots & \theta_{N_T-1}[n]
\end{IEEEeqnarraybox*}\right].
\label{def:Theta[n]}
\end{eqnarray}
Without loss of generality $\theta_0[n]$ is set to zero since the optimal vector is derived from the Frobenius norm operation as in~(\ref{eq:Theta[n]_argmax}). Uniform quantization on each phase steering value is proposed for multiple-input single-output systems in~\cite{Heath98} under the assumption that $\theta_i[n]$'s are statistically independent and uniformly distributed over $[-\pi,~\pi)$. In~\cite{Love03E}, EGT is associated with different types of receiver combining schemes which achieve full diversity order for MIMO systems, and an algorithm is proposed to construct codebooks including predetermined quantized equal gain beamforming vectors. The optimal beamforming vector cannot be determined in a closed-form due to the non-convexity of the problem except for some special cases~\cite{Lee09}. In~\cite{Lee09, Zheng07}, algorithms to find sub-optimal solution for equal gain beamforming vectors are proposed for MIMO systems with arbitrary number of transmit and receive antennas. Construction methods for EGT codebook with PAPC are proposed by using random search algorithms~\cite{Hochwald00} or combinatorial number theory~\cite{Xia05}.

Most of prior works assume block fading channel models, where channel coefficients remain constant for certain amount of time duration and change independently in time to the previous one
\cite{Goldsmith05}. In practice, however, since temporal correlation exists between adjacent channel coefficients for wireless fading channels, one can utilize the previous reported feedback messages as prior information to efficiently encode current feedback message. For Rayleigh fading channels, the correlation coefficient is given
by
\begin{equation}
\rho = E\{h_{ij}[n]h_{ij}^\ast[n-1]\}=J_0(2\pi f_N) \label{def:rho}
\end{equation}
where $h_{ij}[n]$ represents an $(i, j)$-th entry of $\Hb[n]$, $J_0(\cdot)$ is the zero-th order Bessel function of the first kind, and $f_N$ denotes the normalized Doppler frequency (NDF)~\cite{Jakes74}. The NDF is defined as $f_N = f_{D,max}\tau$ where $f_{D,max}$ is the maximum Doppler frequency and $\tau$ is the frame slot duration between $h_{ij}[n]$ and $h_{ij}[n-1]$. Hence, considering temporal correlation property of fading channels, we can adopt time evolution channel generation model as
\begin{equation}
h_{i,j}[n] = \rho h_{i,j}[n-1] + \rho_c u_{i,j}
\label{def:Temporal_model}
\end{equation}
where $\rho_c = \sqrt{1-\rho^2}$, and time evolution term $u_{i,j}$'s are i.i.d. complex Gaussian RV's with zero mean and unit variance.
By exploiting the characteristics of temporal correlation, differential encoding schemes are applied to track time-varying channel coefficient~\cite{Koorapaty05} and to track Givens
parameters for orthogonal frequency division multiplexing (OFDM) systems~\cite{Chin08}--\cite{Roh07}. More recently, the time-varying channel is modeled by finite-state Markov chain and analytical results including information rate, bit rate, and effect of feedback delay are presented with the proposal of an algorithm for compressing CSI feedback~\cite{Huang09}.
In~\cite{Mondal06}--\cite{Samanta05}, codebook switching schemes are proposed to adapt changes in channel distributions. Localized codebook sets based on chordal distance are also utilized to reduce signaling overhead, where feedback index is chosen only from the subset of neighboring codewords from the one at previous time slot~\cite{Sorrentino08}. In~\cite{Banister03a}--\cite{Banister03b}, subspaces of channels are recursively fed back to the transmitter by using a simple gradient approach. Subspace tracking is further
investigated by analyzing the geodesic trajectory connecting two subspaces located in adjacent time instances~\cite{Yang07}. Similar attempts have been made and the performance is evaluated using real data and channel measurements~\cite{Murga09a, Murga09b}.

\begin{figure*}[!t]
\centerline{\subfloat[]{\includegraphics[width=3.5in]{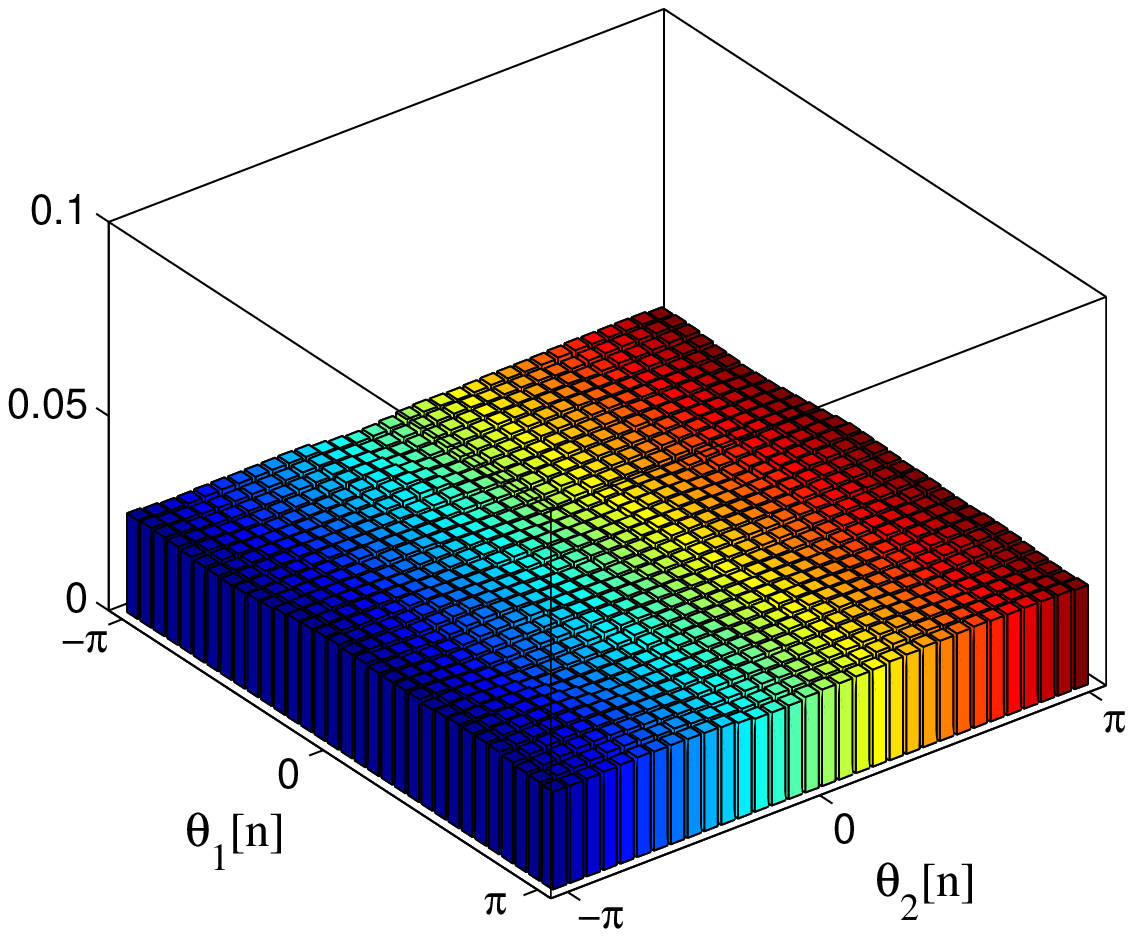}
\label{fig:2D_PDF(a)}}
\hfil
\subfloat[]{\includegraphics[width=3.5in]{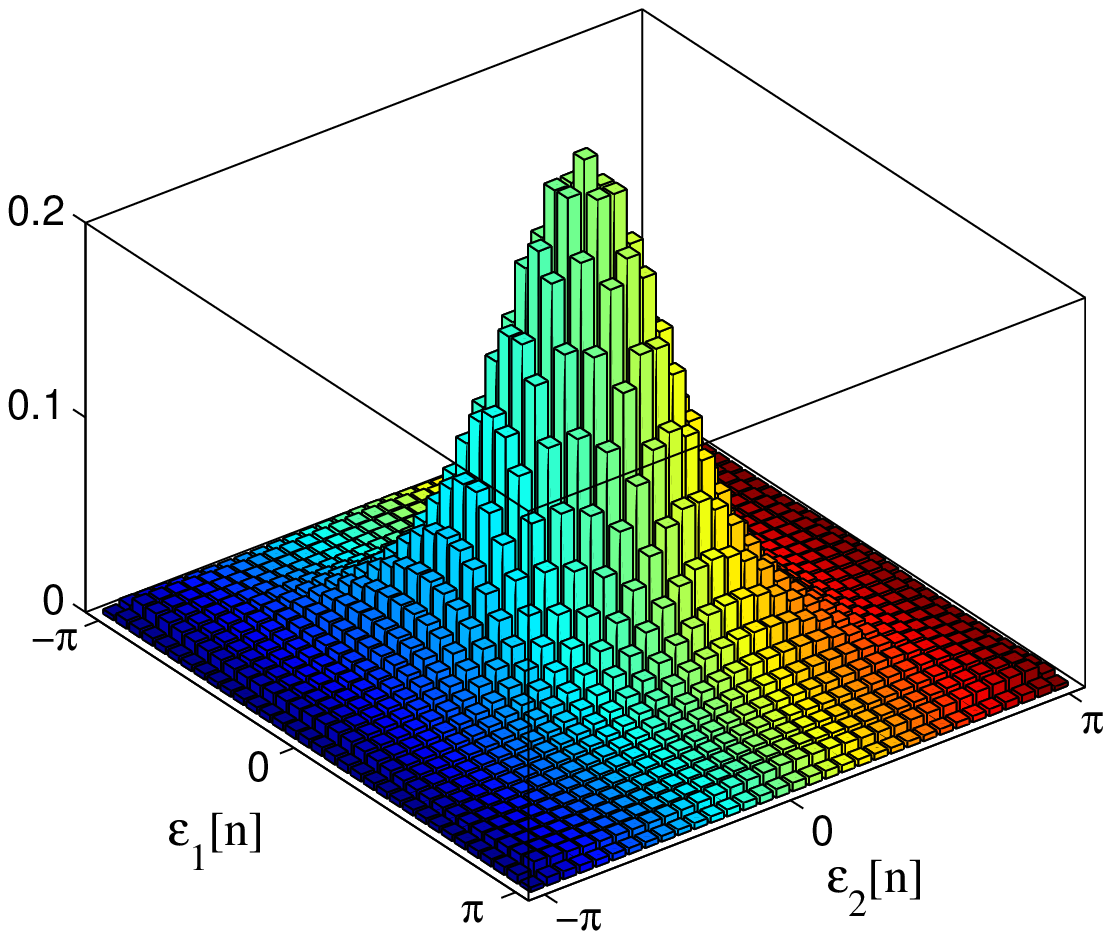}
\label{fig:2D_PDF(b)}}}
\caption{Joint distributions for the elements of (a) phase steering vector, and (b) phase tracking vector ($N_T  = 3$ and $f_N  = 0.1$)}
\label{fig:2D_PDF}
\end{figure*}

In this paper, we consider EGT (which is optimal under PAPC) for MIMO beamforming to obtain diversity and beamforming gains over time-varying fading channels. The optimal phase steering information for each antenna is chosen by (\ref{eq:Theta[n]_argmax}) based on the coefficients in $\Hb[n]$ at the receiver side. Then, the phase steering information is encoded as a feedback messages, and reported to the limited feedback link. Due to the temporal correlation property of the fading channels, to efficiently reduce the amount of feedback overhead, differential encoding can be applied to the phase steering values from adjacent time instances. We define the phase tracking vector as
\begin{equation}
\bm{\epsilon}[n] = \bm{\theta}[n] - \bm{\theta}[n-1] \mod 2\pi.
\label{def:epsilon[n]}
\end{equation}
Unlike the uniformity in the distribution of the phase steering vector~(\ref{def:Theta[n]}) as in Fig.~\ref{fig:2D_PDF}(a), the distribution of the phase tracking vector~(\ref{def:epsilon[n]}) is not uniform as shown in Fig.~\ref{fig:2D_PDF}(b). Temporal correlation causes the non-uniformity in the distribution, and it
can result that the \emph{entropy} of the phase tracking vector is smaller than the phase steering vector. Since the entropy determines the required amount of bits to convey the corresponding information, the required feedback overhead of phase tracking beamforming can be represented by the entropy of the phase tracking vector. From~(\ref{def:epsilon[n]}), we obtain
\begin{eqnarray}
h\!\!\left(\bm{\epsilon}[n]\right) \eqqq h\!\!\left(\bm{\theta}[n] | \bm{\theta}[n-1]\right) \nn\\
                               \eqqq h\!\!\left(\bm{\theta}[n]\right) - I\!\!\left( \bm{\theta}[n]; \bm{\theta}[n-1]\right) \\
                               \leqqq h\!\!\left(\bm{\theta}[n]\right) \nn
\label{eq:h(eps[n])}
\end{eqnarray}
where $h(\cdot)$ and $I(\cdot)$ respectively denote the continuous entropy function and the mutual information (MI) function. Equation (\ref{eq:h(eps[n])}) shows that the continuous entropy is reduced by using differentially encoded feedback messages, and the amount of reduction is equal to the mutual information of $\bm{\theta}[n]$ and $\bm{\theta}[n-1]$. For an extreme example, in the case of static channel environment (channel coefficients are constant in time, i.e., $\rho = 1$), the required number of bits for feedback messages with differential encoding is zero, since the phase steering values are also constant in time. For another extreme case of random channel environment (channel coefficients in adjacent feedback slots are independent RV's, i.e., $\rho = 0$), the required number of bits for feedback messages with differential encoding is identical to the case of coherent encoding, since the phase steering values of a previous time instance give no prior information to the current values. The key objective of this paper is to derive the amount of required feedback messages for both coherent and differential encoding in the function of the temporal correlation parameter between adjacent time instances over time-varying channels.

{\em Organization:} This paper is organized as follows. In Section II, statistical characteristics of feedback messages are investigated. For the phase steering value, exact distribution and entropies are presented. The phase tracking value is represented as a function of RV's whose distributions are derived. However, obtaining the exact distribution and corresponding entropies requires exceedingly complicated probabilistic computations and can not be represented as closed-form expressions. Thus in Section III, upperbounds of entropies of the phase tacking value are derived based on the theory on the maximum entropy distributions. The derived entropy results are transformed to the amount of feedback overhead for two types of feedback messages, and the amount of reduction in feedback overhead over temporally correlated channels is presented in Section IV. In Section V, derived results are jointly compared to the Monte-Carlo simulation results, and suggestions of applications for phase tracking codebook and determination of feedback duration are proposed with some numerical results. Finally, this paper is concluded in Section VI.

{\em Notations and assumptions:} Matrices and vectors are denoted as uppercase and lowercase boldface letters, respectively. The transpose is represented by the superscript ${}^T$ and the Hermitian transposition is designated by the superscript ${}^H$. We use Frobenius norm operator $\|\cdot\|^2$ to obtain the sum of powers of elements in matrices or vectors. $ A \mod B $ denote that $A$ modulor $B$. For a complex variable $x$, $\angle(x)$ represents the angle of $x$; $\Re$, and $\Im$ respectively denote the real, and imaginary component of $x$. The $a$-th order Bessel and modified Bessel functions of the first kind are respectively denoted as $J_a(\cdot)$ and $I_a(\cdot)$. For statistics, $E\{\cdot\}$ stands for expectation and $f_x(\cdot)$ denotes the probability distribution function (PDF) of RV $x$. The distribution equality is written as $X \doteq Y$ when the PDF's of $X$ and $Y$ are identical. Continuous and discrete entropy is defined as $h(\cdot)$ and $H(\cdot)$, respectively.

\section{Statistical Characteristics of Feedback Messages} \label{sec:End_UB}
\subsection{Phase Steering Information}
The objective function in (\ref{eq:Theta[n]_argmax}) is written as
\begin{eqnarray}
\|\Hb[n]\vb[n]\|^2
    \eqqq
        \| \hb_0[n] + \sum_{i=1}^{N_T-1}\hb_i[n]e^{j\theta_i[n]} \|^2 \nn\\
    \eqqq
        \| \hb_{l,c}[n] + \hb_l[n]e^{j\theta_l[n]} \|^2
\label{eq:derivation1}
\end{eqnarray}
where $\hb_l[n]$ denotes $l$-th column vector of channel matrix $\Hb[n]$, and
\begin{eqnarray}
\hb_{l,c}[n] = \hb_0[n] + \sum_{i=1,i \neq l}^{N_T-1}\hb_i[n]e^{j\theta_i[n]}.
\label{eq:derivation2}
\end{eqnarray}
For given $\theta_i[n]$'s with $i \neq l$, the optimal phase steering value of $l$-th antenna is obtained as
\begin{eqnarray}
\theta_l[n] \eqqq
    \underset{x}\argmax
    \| \hb_{l,c}[n] + \hb_l[n]e^{jx} \|^2\nn\\
\eqqq
    \underset{x}\argmax \left\{\| \hb_{l,c}[n] \|^2 + \| \hb_l[n] \|^2
        + 2 \Re \{\hb_{l,c}^H[n]\hb_l[n] e^{j\theta_l[n]} \}\right\}\nn\\
\eqqq
    \angle
    \left(
        \hb_l^H[n] \hb_{l,c}[n]
    \right).
\label{eq:theta_l[n]}
\end{eqnarray}

Following lemma and theorem provide exact distribution and corresponding entropy for phase steering information.

\begin{lemma}
Phase steering values are i.i.d. uniform RV's over $[-\pi,~\pi)$.
\label{Lemma1}
\end{lemma}
\begin{IEEEproof}
Let us define the set of antenna indices $I = \{i\}$ for all integer $i$ satisfying $1\leq i \leq N_T-1$ and $i \neq l$. Prior to the proof, we note that 1) $h(\theta_l[n] | \theta_{i\in I}[n]) \leq h(\theta_l[n])$ always holds due to the fact that conditioning reduces entropy, 2) $h(\theta_l[n]) \leq \log_2 2\pi$ since which is the maximum achievable entropy among all possible PDF's of $\theta_l[n] \in [-\pi,~\pi)$.
\textbf{(Uniformity)} For given $\theta_{i \in I}[n]$'s, $\hb_l[n]$ and $\hb_{l,c}[n]$ become statistically independent complex Gaussian random vectors. Thus $\theta_l[n]$ is uniformly distributed over $[-\pi,~\pi)$ for given $\theta_{i \in I}[n]$'s.
\textbf{(Independency)} Hence we have $h(\theta_l[n] | \theta_{i\in I}[n]) = \log_2 2\pi$ which is known as the maximum continuous entropy for all RV's distributed in $[-\pi,~\pi)$. Thus $h(\theta_l[n] | \theta_{i\in I}[n]) \leq h(\theta_l[n])$ holds with equality, which represents the independency between $\theta_l[n]$ and $\theta_{i\in I}[n]$'s.
\end{IEEEproof}

\begin{theorem}
Continuous entropy of the phase steering value and the phase steering vector respectively become $h\!\!\left({\theta_l}[n]\right) = \log_2 2\pi$ and $h\!\!\left(\bm{\theta}[n]\right) = (N_T - 1)\log_2 2\pi$.
\label{Theorem1}
\end{theorem}
\begin{IEEEproof}
From the uniformity and independency among phase steering values shown in {\em Lemma \ref{Lemma1}}, the entropy of the phase steering value becomes $h(\theta_l[n]) = \log_2 2\pi$ and the entropy of the phase steering vector becomes $h\!\!\left(\bm{\theta}[n]\right) = \sum_{l=1}^{N_T - 1}h(\theta_l[n]) = (N_T - 1)\log_2 2\pi$.
\end{IEEEproof}

\subsection{Phase Tracking Information}
Considering the channel generation model in (\ref{def:Temporal_model}), we let
\begin{eqnarray}
\hb_l[n] \eqqq \rho\hb_l[n-1] + \rho_c\ub_l \\
\hb_{l,c}[n] \eqqq \rho\hb_{l,c}[n-1] + \rho_c\ub_{l,c}
\label{eq:temporal_corr}
\end{eqnarray}
where the elements in $\ub_l$ are i.i.d. complex Gaussian RV's with zero mean and unit variance and the elements in $\ub_{l,c}$ are identical but with variance of $N_T-1$ (the variance of elements in $\hb_{l,c}[n-1]$). Then, (\ref{eq:theta_l[n]}) becomes
\begin{eqnarray}
\theta_l[n] = \angle \!
\left(\! \rho\hb_l^H[n-1] \!+\! \rho_c\ub_l^H \!\right)\!\!
\left(\! \rho\hb_{l,c}[n-1] \!+\! \rho_c\ub_{l,c} \!\right).
\label{eq:theta_l[n]2}
\end{eqnarray}
Let us further define
\begin{eqnarray}
\begin{IEEEeqnarraybox*}[][c]{,l/l,}
    \alpha_1 = \hb_l^H[n-1]\hb_{l,c}[n-1],  &   \alpha_2 = \hb_l^H[n-1]\ub_{l,c}, \\
    \alpha_3 = \ub_l^H\hb_{l,c}[n-1],  &   \alpha_4 = \ub_l^H\ub_{l,c},
\end{IEEEeqnarraybox*}
\label{def:alpha's}
\end{eqnarray}
where $\alpha_i$'s are the sums of products of two Gaussian RV's, which are assumed to be i.i.d. zero mean complex Gaussian RV's from the central limit theorem for large $N_R$ (the discrepancy arising from this assumption is shown to be minimal even for $N_R = 2$ using numerical results). We can rewrite (\ref{eq:theta_l[n]2}) as
\begin{eqnarray}
\theta_l[n]
\eqqq
\angle \!\! \left(\rho^2\alpha_1 + \rho_c\rho(\alpha_2 + \alpha_3) + \rho_c^2 \alpha_4 \right)\nn\\
\!\!\!\!&\underset{(a)}{=}&\!\!\!\!\
\angle \!\! \left( e^{j\theta_l[n-1]} + \frac{\rho_c}{\rho|\alpha_1|}(\alpha_2+\alpha_3) + \frac{\rho_c^2}{\rho^2|\alpha_1|}\alpha_4 \right) \nn\\
\eqqq
\angle \!\! \left( e^{j\theta_l[n-1]} + \gamma e^{j \phi} \right)
\label{eq:theta_l[n]3}
\end{eqnarray}
where (a) follows from the fact that $\angle\alpha_1$ represents $\theta_l[n-1]$ as shown in~(\ref{eq:theta_l[n]}); $\gamma$ and $\phi$ are respectively defined as the magnitude and the phase of $\frac{\rho_c}{\rho|\alpha_1|}(\alpha_2+\alpha_3) + \frac{\rho_c^2}{\rho^2|\alpha_1|}\alpha_4$. Considering relation in~(\ref{def:epsilon[n]}), the tracking value of $l$-th antenna (for notational convenience, $\epsilon_l[n]$ is denoted by $\epsilon$) is given as
\begin{eqnarray}
\epsilon
\eqqq
\angle \! \left\{\!\left(\!\! e^{j\theta_l[n-1]} + \gamma e^{j \phi} \right)\!\!e^{-j \theta_l[n-1]}\! \right\}\nn \\
\eqqq
\angle \! \left( 1 + \gamma e^{j \psi} \right)
\label{def:epsilon}
\end{eqnarray}
where $\psi = \phi - \theta_l[n-1]$. We define $\gamma e^{j \psi}$ as the {\em feedback overhead generator} (FOG).

\begin{proposition}
The phase of FOG is a uniformly distributed RV over $[-\pi,~\pi)$.
\label{Proposition:Phase_FOG}
\end{proposition}
\begin{IEEEproof}
Under the assumption that $\alpha_i$'s are i.i.d. zero mean complex Gaussian RV's, the phase $\phi$ in (\ref{eq:theta_l[n]3}) is uniformly distributed over $[-\pi,~\pi)$ and statistically independent to $\angle\alpha_1$ since it is formed by a weighted sum of $\alpha_2, \alpha_3$, and $\alpha_4$. Thus the phase of FOG $\psi = \phi - \theta_l[n-1]$ is also uniformly distributed over $[-\pi,~\pi)$ due to the fact that $\angle\alpha_1 = \theta_l[n-1]$.
\end{IEEEproof}

\begin{proposition}
The distribution of the magnitude of FOG is given by
\begin{eqnarray}
f_{\gamma}(x) = \frac{2 k x}{(1+k x^2)^2} \nn
\end{eqnarray}
where $k = \frac{\rho^4}{1 - \rho^4}$.
\label{Proposition:Mag_FOG}
\end{proposition}
\begin{IEEEproof}
Under the assumption that $\alpha_i$'s are i.i.d. zero mean complex Gaussian RV's, we have the following distribution equality from (\ref{eq:theta_l[n]3})
\begin{eqnarray}
\gamma e^{j\phi} \doteq \sqrt{2\left(\frac{\rho_c}{\rho}\right)^2 + \left(\frac{\rho_c}{\rho}\right)^4} \frac{\alpha_5}{|\alpha_1|}  = \sqrt{\frac{1- \rho^4}{\rho^4}} \frac{\alpha_5}{|\alpha_1|}
\label{eq:Distribution_equality}
\end{eqnarray}
where $X \doteq Y$ denotes that the PDF's of $X$ and $Y$ are identical; $\alpha_5$ is a newly defined zero mean complex Gaussian RV which is i.i.d. to $\alpha_{i\in\{1,\ldots,4\}}$. Thus $\gamma^2$ is the ratio of two exponential RV's. Since the ratio of two i.i.d. exponential RV's has a probability density function (PDF) of $(1+x)^{-2}$, the PDF of $\gamma^2$ is derived as $f_{\gamma^2}(x) =  {k}{(1+k x)^{-2}}$ where $k = \frac{\rho^4}{1 - \rho^4}$. Hence the PDF of $\gamma$, the ratio of two Rayleigh RV's, is immediately obtained from $f_{\gamma^2}(x)$ as $f_{\gamma}(x) = {2 k x}{(1+k x^2)^{-2}}$.
\end{IEEEproof}
Obtaining exact distribution and corresponding entropy of the phase tracking value seems mathematically intractable. By exploiting the statistical characteristics given in {\em Propositions 1} and {\em 2}, we derive analytical upperbounds on the entropy of the phase tracking value in the following section.

\section{Upperbounds on Continuous Entropy of Phase Tracking Information}

Derivation of continuous entropy of the phase tracking value requires corresponding PDF which is not known. In order to present upperbounds on continuous entropy, theory on maximum entropy distributions is firstly reviewed with following three representative cases.
\begin{itemize}[\IEEEsetlabelwidth{Z}]
\item {\em Gaussian Distribution:} For given variance $\sigma^2$, the continuous entropy is maximized when a RV $G$ has the Gaussian PDF
\begin{eqnarray}
f_{G}(x) = \frac{1}{\sqrt{2\pi\sigma^2}} e^{-\frac{(x - \mu)^2}{2\sigma^2}} \nn
\end{eqnarray}
where $\mu$ and $\sigma^2$ respectively denote the mean and variance of the RV $G$. The continuous entropy of the Gaussian PDF is given by
\begin{eqnarray}
h(G) = \log_2 \sqrt{2\pi e \sigma^2}.\nn
\end{eqnarray}
\item {\em Uniform Distribution:} For given interval $[a,~b]$, the continuous entropy is maximized when a RV $U$ follows the uniform PDF
\begin{eqnarray}
f_{U}(x) = \frac{1}{|b-a|} \nn
\end{eqnarray}
for $a \leq x \leq b$, and zero otherwise. The continuous entropy of the uniform PDF is known as
\begin{eqnarray}
h(U) = \log_2 |b-a|.\nn
\end{eqnarray}
\item {\em von-Mises Distribution:} For a circular RV with given concentration parameter $\kappa$, the continuous entropy is maximized when a RV $V$ follows the von-Mises PDF
\begin{eqnarray}
f_{V}(x; \bar{\mu}, \kappa) = \frac{1}{2\pi I_0(\kappa)} e^{\kappa \cos(x-\bar{\mu})}
\label{PDF:von-Mises}
\end{eqnarray}
where $\bar{\mu}$ is a mean direction of $V$. Note that $\bar{\mu} \in [-\pi,~\pi)$ and $\kappa \geq 0$. The corresponding mean resultant length (MRL) is written as
\begin{eqnarray}
\bar{R} = \frac{I_1(\kappa)}{I_0(\kappa)} \triangleq A(\kappa).
\label{eq:Def_R_bar_CG}
\end{eqnarray}
For the von-Mises PDF, the continuous entropy is given by
\begin{eqnarray}
h(V) = -\kappa \bar{R} \log_2 e + \log_2 (2\pi I_0(\kappa)).
\label{def:von-Mises entropy}
\end{eqnarray}
\end{itemize}

Now we are ready to utilize one of above PDFs to derive upperbounds of the exact entropy for the phase tracking value based on the maximum entropy theory. In Subsection A, we first use the Gaussian PDF to obtain upperbounds in mathematically compact expressions. In Subsection B, we derive tighter upperbounds by using the von-Mises PDF, followed by the asymptotic analysis on the derived results to provide insights on the behavior of the phase tracking value in Subsection C.

\subsection{Upperbound Using Gaussian PDF}

From the theory on maximum entropy distributions, the upperbound on continuous entropy can be derived based on the assumption that the phase tracking value has the Gaussian PDF with given variance. The variance of the phase tracking value is calculated from
\begin{eqnarray}
\sigma_{\epsilon}^2 = \int_{0}^{\infty} \sigma_{\epsilon}^2 (\gamma = y) f_{\gamma}(y)dy
\label{eq:var_eps}
\end{eqnarray}
where $\sigma_{\epsilon}^2 (\gamma)$ is the conditional variance of $\epsilon$ for given $\gamma$, and the PDF of the magnitude of FOG $f_{\gamma}(y)$ is given in {\em Proposition \ref{Proposition:Mag_FOG}}.

\begin{theorem}
The continuous entropy of the phase tracking value is upperbounded by $\log_2 \sqrt{2\pi e \sigma_{\epsilon}^2}  {\triangleq} h_G(\epsilon;\sigma_{\epsilon}^2)$.
\label{Theorem:PTV_Gaussian1}
\end{theorem}
\begin{IEEEproof}
It is an immediate consequence of the fact that the Gaussian PDF has the maximum continuous entropy for given variance.
\end{IEEEproof}

However the exact calculation of (\ref{eq:var_eps}) requires exceedingly complicated probabilistic computations, and can not be written in a closed-form expression as well as the derived upperbound in {\em Theorem \ref{Theorem:PTV_Gaussian1}}. Note that the Gaussian entropy $\log_2 \sqrt{2\pi e \sigma^2}$ is monotonic increasing function with respect to the corresponding variance $\sigma^2$. Hence, instead of using the exact value of $\sigma_{\epsilon}^2 $, we use an upperbound of it which can be written in a closed-form.

\begin{lemma}
Conditional variance of $\epsilon$ for given $\gamma$ is upperbounded as
\begin{eqnarray}
\sigma_{\epsilon}^2(\gamma)
    \leqqq \left\{\begin{IEEEeqnarraybox}[\relax][c]{l's}
            \frac{1}{\sqrt{1-\gamma^2}}-1,&       for $0 \leq \gamma \leq 1$\\
            \pi^2/3, &     otherwise.%
          \end{IEEEeqnarraybox}\right.\nn
\end{eqnarray}
\label{Lemma:UB_var_Gaussian}
\end{lemma}
\begin{IEEEproof}
From (\ref{def:epsilon}), $|\epsilon|$ is equal to $|\psi|$ for $\gamma = \infty$ and less than $|\psi|$ otherwise, and thus $\sigma_{\epsilon}^2 \leq \sigma_{\psi}^2 = \frac{\pi^2}{3}$ holds. For $\gamma \leq 1$, the phase tracking value can be written as
\begin{eqnarray}
\epsilon = \tan^{-1}\left(\frac{\gamma \sin\psi}{1+\gamma\cos\psi}\right).
\end{eqnarray}
Since it is known that $|\tan^{-1}x| \leq |x|$, the variance of ${\epsilon}$ for given $\gamma$ is upperbounded by
\begin{eqnarray}
\sigma_{\epsilon}^2(\gamma)
    \leqqq
        \frac{1}{2\pi}\int_{-\pi}^{\pi} \left(\frac{\gamma \sin\psi}{1+\gamma\cos\psi}\right)^2 d\psi \nn\\
    \eqqq  \frac{1}{\sqrt{1-\gamma^2}}-1
\end{eqnarray}
for $\gamma \leq 1$.
\end{IEEEproof}

By using {\em Lemma \ref{Lemma:UB_var_Gaussian}}, the upperbound on $\sigma_{\epsilon}^2$ is derived as
\begin{eqnarray}
\sigma_{\epsilon}^2 \leqqq \int_{0}^{1} \left(\frac{1}{\sqrt{1-\gamma^2}}-1\right) f_{\gamma}(y)dy + \frac{\pi^2}{3} \int_{1}^{\infty} f_{\gamma}(y)dy \nn\\
    \eqqq \sqrt{\frac{k}{(1+k)^3}} \sinh^{-1}\!\!\sqrt{k} + \frac{\pi^2}{3+3k} \triangleq \sigma_u^2,
\label{eq:UB_var_eps}
\end{eqnarray}
and the corresponding closed-form upperbound based on the Gaussian PDF is obtained as follows.

\begin{theorem}
The continuous entropy of the phase tracking value $h(\epsilon)$ and its upperbound $h_G(\epsilon;\sigma_{\epsilon}^2)$ are further upperbounded by $\log_2 \sqrt{2 \pi e \sigma_u^2}  {\triangleq} h_G(\epsilon;\sigma_{u}^2)$.
\label{Theorem:PTV_Gaussian2}
\end{theorem}
\begin{IEEEproof}
In {\em Theorem \ref{Theorem:PTV_Gaussian1}}, it is proved that $\log_2 \sqrt{2 \pi e \sigma_{\epsilon}^2}$ upperbounds the continuous entropy of the phase tracking value. Since $\log_2 (\cdot)$ is a monotonic increasing function and $\sigma_{\epsilon}^2 \leq \sigma_u^2$, we have the desired inequality $\log_2 \sqrt{2 \pi e \sigma_{\epsilon}^2}  \leq \log_2 \sqrt{2 \pi e \sigma_u^2}$. \end{IEEEproof}

\begin{figure*}[!t]
\centerline{\subfloat[]{\includegraphics[width=3.5in]{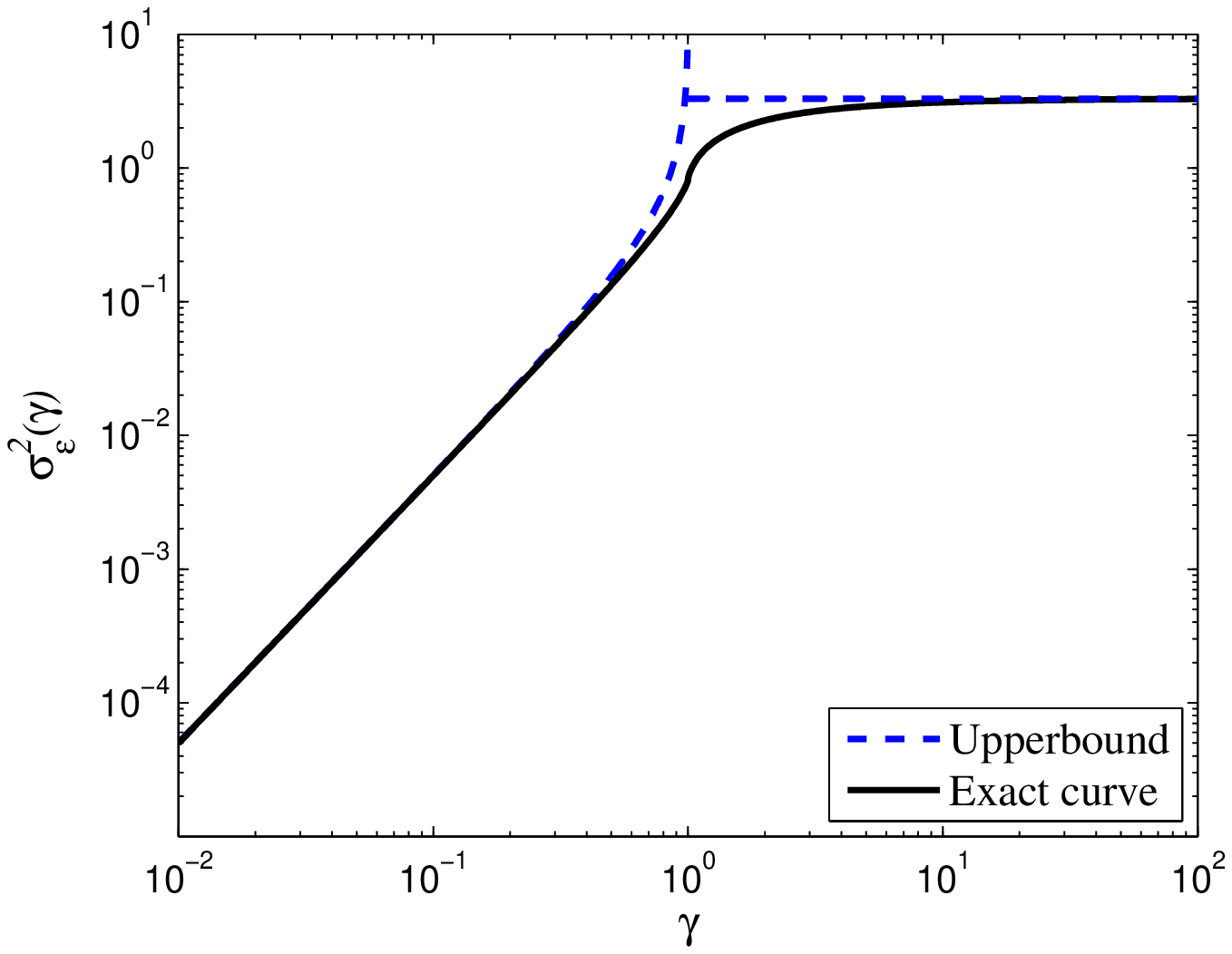}
\label{fig:Cond_var}}
\hfil
\subfloat[]{\includegraphics[width=3.5in]{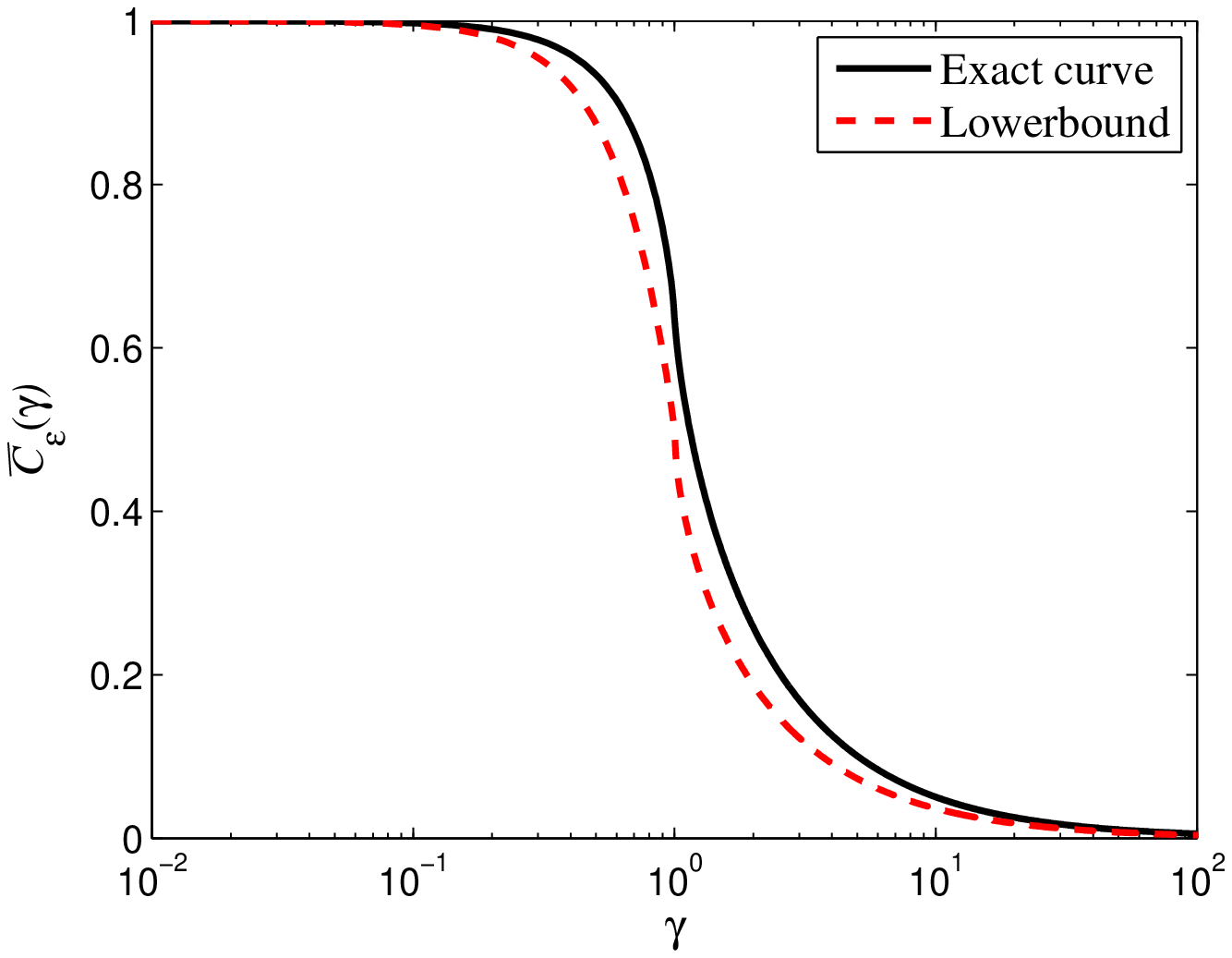}
\label{fig:C_b}}}
\caption{(a) Conditional variance of $\epsilon$ for given $\gamma$ obtained from Monte-Carlo simulation is compared with its corresponding upperbound. (b) Comparison of exact curve for $\bar{C}_{\epsilon}(\gamma)$ in (\ref{eq:Exact_C_gamma}) and its lowerbounds: $\bar{C}_{L_1}(\gamma)$ for $0 \leq \gamma < 1$ in (\ref{eq:LB1_C_gamma}) and $\bar{C}_{L_2}(\gamma)$ for $\gamma \geq 1$ in (\ref{eq:LB2_C_gamma}).}
\label{fig:Variance and MRL}
\end{figure*}

\subsection{Upperbound Using Von-Mises PDF}
Note that the phase tracking value $\epsilon$ is a RV distributed over angular domain $[-\pi,~\pi)$. Hence, we can apply directional statistics to analyze the characteristics of $\epsilon$. Let $x$ be a directional data which has a value in the angular domain $[-\pi,~\pi)$. We further define $\bar{C}_x = E\{\cos x\}$ and $\bar{S}_x = E\{\sin x\}$. Then, the MRL and the mean direction are respectively defined as $\bar{R}_x = \sqrt{\bar{C}_x^2 + \bar{S}_x^2 }$ and $\bar{\mu}_x = \arctan\left({\bar{S}_x}/{\bar{C}_x}\right)$ to characterize the angular data $x$~\cite{Mardia72}. From the definition of the phase tracking value in (\ref{def:epsilon}), let
\begin{eqnarray}
\nu e^{j \epsilon}
\eqqq 1 + \gamma e{j\psi} \nn\\
\eqqq 1+\gamma\cos\psi + j\gamma\sin\psi, \nn
\end{eqnarray}
then we can obtain $\nu = \sqrt{1 + \gamma^2 + 2\gamma \cos \psi}$. Thus we have $\cos\epsilon = \frac{1+\gamma\cos\psi}{\sqrt{1 + \gamma^2 + 2\gamma \cos \psi}}$ and $\sin\epsilon = \frac{\gamma\sin\psi}{\sqrt{1 + \gamma^2 + 2\gamma \cos \psi}}$. Taking expectation of $\sin\epsilon$, we obtain
\begin{eqnarray}
\bar{S}_{\epsilon}
    \eqqq E\{\sin \epsilon \}
        = E\{ \frac{\gamma \sin\psi}{\sqrt{1 + \gamma^2 + 2\gamma \cos \psi}} \} \nn \\
    \eqqq \frac{1}{2\pi} \int_{-\pi}^{\pi} \frac{\gamma \sin\psi}{\sqrt{1 + \gamma^2 + 2\gamma \cos \psi}} d\psi \nn\\
    \eqqq 0
\label{eq:S_bar}
\end{eqnarray}
where the last equality comes from the fact that
\begin{eqnarray}
\left.\frac{\gamma \sin\psi}{\sqrt{1 + \gamma^2 + 2\gamma \cos \psi}}\right|_{\psi = x} = -\left.\frac{\gamma \sin\psi}{\sqrt{1 + \gamma^2 + 2\gamma \cos \psi}}\right|_{\psi = -x}. \nn
\end{eqnarray}
Hence the MRL and the mean direction for $\epsilon$ are respectively given as $\bar{R}_{\epsilon} = |\bar{C}_{\epsilon}|$, and $\bar{\mu}_{\epsilon} = 0$.

To derive $\bar{C}_{\epsilon}$, we first obtain the conditional expectation for given $\gamma$ as
\begin{eqnarray}
\bar{C}_{\epsilon}(\gamma)
    \eqqq E\{\cos \epsilon | \gamma\}
        = E\left\{ \left.\frac{1 + \gamma \cos \psi}{\sqrt{1 + \gamma^2 + 2 \gamma \cos \psi}} \right| \gamma \right\} \nn \\
    \eqqq \frac{1}{2\pi} \int_{-\pi}^{\pi} \frac{1 + \gamma \cos\psi}{\sqrt{1 + \gamma^2 + 2\gamma \cos \psi}} d\psi \nn\\
    \eqqq \left\{\begin{IEEEeqnarraybox}[\relax][c]{l's}
            g(\gamma),&       for $0 \leq \gamma < 1$\\
            -g(\gamma), &     otherwise%
          \end{IEEEeqnarraybox}\right.
\label{eq:Exact_C_gamma}
\end{eqnarray}
where $g(\gamma)$ is defined as
\begin{eqnarray}
g(\gamma) \eqqq \frac{1}{\pi} \left\{ (\gamma - 1)E_E\left( \frac{-4\gamma}{(\gamma - 1)^2} \right)
          + (\gamma + 1)E_K\left( \frac{-4\gamma}{(\gamma - 1)^2} \right) \right\}
\end{eqnarray}
with the aid of the complete elliptic integral of the first kind function $E_K$ and the second kind function $E_E$ \cite{Abramowitz64}. The expectation over $\gamma$ requires to calculate
\begin{eqnarray}
\bar{C}_{\epsilon} = \int_0^{\infty} \bar{C}_{\epsilon}(x) f_{\gamma}(x)dx.
\label{eq:Exp_C_gamma}
\end{eqnarray}
For the following theorem, we assume that $\bar{C}_{\epsilon} \geq 0$, and thus $\bar{R}_{\epsilon} = \bar{C}_{\epsilon}$ is satisfied (It will be proved in {\em Proposition \ref{Proposition:R_bar LB}}).

\begin{theorem}
Let the inverse function of (\ref{eq:Def_R_bar_CG}) as $A^{-1}(\bar{R}_{\epsilon}) = \kappa_{\epsilon}$, then the continuous entropy of the phase tracking value is upperbounded by
$- \kappa_{\epsilon} \bar{R}_{\epsilon}\log_2 e + \log_2 (2\pi I_0({\kappa}_{\epsilon}))  {\triangleq}  h_V(\epsilon;\bar{R}_{\epsilon})$.
\label{Theorem:PTV_von-Mises1}
\end{theorem}
\begin{IEEEproof}
It is an immediate consequence of the fact that von-Mises PDF has the maximum continuous entropy of angular data for given concentration parameter.
\end{IEEEproof}

However, (\ref{eq:Exp_C_gamma}) can not be represented in a closed-form expression. Note that the continuous entropy of the von-Mises PDF is inversely proportional to the corresponding MRL $\bar{R}_{\epsilon}$ (See Appendix A). Thus we derive a closed-form upperbound on the continuous entropy of the phase tracking value by considering a closed-form lowerbound of $\bar{R}_{\epsilon}$, equivalently a lowerbound of $\bar{C}_{\epsilon}$ due to $\bar{R}_{\epsilon} = \bar{C}_{\epsilon}$ (It will be proved in {\em Proposition \ref{Proposition:R_bar LB}}).

\begin{lemma}
The conditional expectation $\bar{C}_{\epsilon}(\gamma)$ is lowerbounded as
\begin{eqnarray}
\bar{C}_{\epsilon}(\gamma) \leq
\left\{\begin{IEEEeqnarraybox}[\relax][c]{l's}
            1 - \frac{\gamma^2}{2} {~\triangleq~} \bar{C}_{L_1}(\gamma),&       for $0 \leq \gamma < 1$\\
            \frac{1}{(\gamma + 1)} + \frac{2\tan^{-1}\sqrt{\gamma^2 - 1}}{\pi(\gamma + 1)(\gamma - 1)}
        + \frac{-2}{\pi\sqrt{\gamma^2 - 1}} {~\triangleq~} \bar{C}_{L_2}(\gamma), &     otherwise.%
          \end{IEEEeqnarraybox}\right.
\end{eqnarray}
\label{Lemma:LB_MRL}
\end{lemma}
\begin{IEEEproof}
See Appendix B.
\end{IEEEproof}

By using the lowerbounds $\bar{C}_{L_1}(\gamma)$ and $\bar{C}_{L_2}(\gamma)$, we obtain the lowerbound of $\bar{C}_{\epsilon}$ as
\begin{eqnarray}
\bar{C}_{\epsilon}
    \!\!\!\!&{\geq}&\!\!\!\! \underset{\triangleq c_1(k)}{\underbrace{\int_0^{1} \bar{C}_{L_1}(x) f_{\gamma}(x)dx}} + \underset{\triangleq c_2(k)}{\underbrace{\int_1^{\infty} \bar{C}_{L_2}(x) f_{\gamma}(x)dx}}
\label{eq:Exp_LB_C}
\end{eqnarray}
where
\begin{eqnarray}
c_1(k) \eqqq 1 - \frac{1}{2(1+k)} - \frac{\ln(1+k)}{2k} \label{eq:c_1(k)}\\
c_2(k) \eqqq \frac{\sqrt{k}(1-k)\left(\pi - 2\tan^{-1}\sqrt{k} \right)}{2(1+k)^2}
            + \frac{k}{(1+k)^2}\ln\left(\frac{4k}{1+k}\right) \nn\\
           &&-\frac{k}{(1+k)^2} \left\{ 2 \ln\left(\sqrt{k(1+k)}-k\right)
                + k + 1 - \sqrt{k(1+k)}\right\} - \sqrt{\frac{k}{(1+k)^3}}.\label{eq:c_2(k)}
\end{eqnarray}
\begin{proposition}
(a) $\bar{C}_{\epsilon}$ has a non-negative value, (b) $\bar{R}_{\epsilon} = \bar{C}_{\epsilon}$ is satisfied, and (c) the MRL $\bar{R}_{\epsilon}$ is lowerbounded by $c_1(k) + c_2(k) {\triangleq} \bar{R}_L$.
\label{Proposition:R_bar LB}
\end{proposition}
\begin{IEEEproof}
(a) The integration values in (\ref{eq:Exp_LB_C}) are non-negative, i.e., $c_1(k) \geq 0$ and $c_2(k) \geq 0$ (See Appendix C). Thus, from the inequality in (\ref{eq:Exp_LB_C}), we have $ \bar{C}_{\epsilon} \geq c_1(k) + c_2(k) {\geq} 0$. (b) Since $\bar{S}_{\epsilon} = 0$ as shown in (\ref{eq:S_bar}), the MRL is derived as $\bar{R}_{\epsilon} = \sqrt{\bar{C}_{\epsilon}^2 + \bar{S}_{\epsilon}^2 } = \bar{C}_{\epsilon}$. (c) Hence we have the lowerbound on the MRL as $\bar{R}_{\epsilon} \geq c_1(k) + c_2(k)$ due to the inequality in (\ref{eq:Exp_LB_C}) and the relation of $\bar{R}_{\epsilon} = \bar{C}_{\epsilon}$.
\end{IEEEproof}
\begin{theorem}
Let $A^{-1}(\bar{R}_{L}) = \kappa_{L}$, then the continuous entropy of the phase tracking value $h(\epsilon)$ and its upperbound $h(\epsilon;\bar{R}_{\epsilon})$ are further upperbounded by $-{\kappa}_{L} \bar{R}_{L}\log_2 e + \log_2 (2\pi I_0({\kappa}_{L})) {\triangleq}  h_V(\epsilon;\bar{R}_{L})$.
\label{Theorem:PTV_von-Mises2}
\end{theorem}
\begin{IEEEproof}
The continuous entropy of the von-Mises PDF is inversely proportional to the corresponding MRL (See Appendix A). Proof completes by noting that $\bar{R}_{\epsilon} \geq \bar{R}_L$ as proved in {\em Proposition \ref{Proposition:R_bar LB}}.
\end{IEEEproof}

\subsection{Asymptotic Behavior of the Derived Upperbounds}
Since four upperbounds on the continuous entropy of the phase tracking value have been derived based on the assumption that the corresponding PDF is the Gaussian PDF or the von-Mises PDF. Two of them in {\em Theorem \ref{Theorem:PTV_Gaussian2}} and {\em Theorem \ref{Theorem:PTV_von-Mises2}} are closed-form expressions though which are respectively looser than the corresponding numerical bounds in {\em Theorem \ref{Theorem:PTV_Gaussian1}} and {\em Theorem \ref{Theorem:PTV_von-Mises1}}. In this subsection, we present the asymptotic behavior of the exact continuous entropy of the phase tracking value along with the derived closed-form upperbounds. Then, the asymptotic tightness are discussed by comparing the derived closed-form upperbounds and the exact one.

\subsubsection{Exact Entropy}
Since $\lim_{\rho \rightarrow 1} \gamma = 0$ and $\lim_{\rho \rightarrow 0} \gamma = \infty$ from the distribution equality in (\ref{eq:Distribution_equality}), the phase tracking value becomes
\begin{eqnarray}
\lim_{\rho \rightarrow 1}\epsilon \eqqq \angle(1+\gamma e^{j\psi}) = 0, \label{eq:asymp_exact1} \\
\lim_{\rho \rightarrow 0}\epsilon \eqqq \angle(1+\gamma e^{j\psi}) = \psi \label{eq:asymp_exact2}.
\end{eqnarray}
Thus we have
\begin{eqnarray}
\lim_{\rho \rightarrow 1} h(\epsilon) \eqqq -\infty
\end{eqnarray}
the continuous entropy of a constant value, and
\begin{eqnarray}
\lim_{\rho \rightarrow 0} h(\epsilon) \eqqq \log_2 2\pi
\end{eqnarray}
the continuous entropy of a uniform RV distributed over $[-\pi,~\pi)$.

\subsubsection{Gaussian Upperbound in Theorem \ref{Theorem:PTV_Gaussian2}}
The upperbound on the variance of the phase tracking value in (\ref{eq:UB_var_eps}) is asymptotically written as
\begin{eqnarray}
\lim_{\rho \rightarrow 1} \sigma_u^2
    \!\!\!\!&\underset{(a)}{=}&\!\!\!\! \lim_{k \rightarrow \infty}\sigma_u^2
    =
        \lim_{k \rightarrow \infty} \frac{\pi^2}{3+3k} + \lim_{k \rightarrow \infty} \sqrt{\frac{k}{(1+k)^3}} \sinh^{-1}\!\!\sqrt{k} = 0, \label{eq:asymp_Gauss1}\\
\lim_{\rho \rightarrow 0} \sigma_u^2
    \!\!\!\!&\underset{(a)}{=}&\!\!\!\! \lim_{k \rightarrow 0}\sigma_u^2
    =
        \lim_{k \rightarrow 0} \frac{\pi^2}{3+3k} + \lim_{k \rightarrow 0} \sqrt{\frac{k}{(1+k)^3}} \sinh^{-1}\!\!\sqrt{k} = \frac{\pi^2}{3} \label{eq:asymp_Gauss2},
\end{eqnarray}
where (a) follows from the relation $k=\frac{\rho^4}{1-\rho^4}$. By the definition of $h_G(\epsilon; \sigma_u^2) = \log_2 \sqrt{2\pi e \sigma_u^2}$, we have
\begin{eqnarray}
\lim_{\rho \rightarrow 1} h_G(\epsilon; \sigma_u^2) = -\infty, \\
\lim_{\rho \rightarrow 0} h_G(\epsilon; \sigma_u^2) = \frac{1}{2} \log_2 \frac{2 e \pi^3}{3}.
\end{eqnarray}

\subsubsection{von-Mises Upperbound in Theorem \ref{Theorem:PTV_von-Mises2}}

From (\ref{eq:c_1(k)})--(\ref{eq:c_2(k)}), we recognize that
\begin{eqnarray}
\lim_{k \rightarrow \infty} c_1(k) = 1, && \lim_{k \rightarrow 0} c_1(k) = 0, \\
\lim_{k \rightarrow \infty} c_2(k) = 0, && \lim_{k \rightarrow 0} c_2(k) = 0.
\end{eqnarray}
Thus the lowerbound on the MRL is asymptotically given as
\begin{eqnarray}
\lim_{\rho \rightarrow 1} \bar{R}_L
    \!\!\!\!&\underset{(a)}{=}&\!\!\!\! \lim_{k \rightarrow \infty} \left\{c_1(k) + c_1(k)\right\} = 1, \label{eq:asymp_von1} \\
\lim_{\rho \rightarrow 0} \bar{R}_L
    \!\!\!\!&\underset{(a)}{=}&\!\!\!\! \lim_{k \rightarrow 0} \left\{c_1(k) + c_1(k)\right\} = 0, \label{eq:asymp_von2}
\end{eqnarray}
where (a) follows from $k=\frac{\rho^4}{1-\rho^4}$. Note that $\kappa_L = A^{-1}(\bar{R}_L)$ becomes infinity as $\bar{R}_L$ goes to one. According to \cite{Mardia72}, the von-Mises PDF has a limiting behavior for large $\kappa_L$ as the Gaussian PDF
\begin{eqnarray}
\lim_{\kappa_L \rightarrow \infty} f_{V}(x; \bar{\mu}, \kappa_L) =
    \frac{1}{\sqrt{2\pi\tilde{\sigma}^2}} e^{-\frac{(x - \bar{\mu})^2}{2\tilde{\sigma}^2}}
\end{eqnarray}
where $\tilde{\sigma}^2 = 1/\kappa_L$ which approaches to zero as $\kappa_L$ goes to infinity. The corresponding continuous entropy is written as
\begin{eqnarray}
\lim_{\rho \rightarrow 1} h_V(\epsilon; \bar{R}_L)
    \eqqq \lim_{\bar{R}_L \rightarrow 1} h_V(\epsilon; \bar{R}_L) = \lim_{\kappa_L \rightarrow \infty} \log_2 \sqrt{\frac{2 \pi e}{\kappa_L}} = -\infty.
\end{eqnarray}
On the other hand, $\kappa_L = A^{-1}(\bar{R}_L)$ becomes zero as $\bar{R}_L$ goes to zero. Also from \cite{Mardia72}, the von-Mises PDF has a limiting behavior for small $\kappa_L$ as the uniform PDF
\begin{eqnarray}
\lim_{\kappa_L \rightarrow 0} f_{V}(x; \bar{\mu}, \kappa) =
    \frac{1}{2\pi}
\end{eqnarray}
for $x \in [-\pi,~\pi)$ and zero otherwise. The corresponding continuous entropy is given by
\begin{eqnarray}
\lim_{\rho \rightarrow 0} h_V(\epsilon; \bar{R}_L)
    \eqqq \lim_{\bar{R}_L \rightarrow 0} h_V(\epsilon; \bar{R}_L) = \lim_{\kappa_L \rightarrow 0} h_V(\epsilon; \bar{R}_L) = \log_2 2 \pi.
\end{eqnarray}

\subsubsection{Asymptotic tightness of the Derived Upperbounds}
As $\rho$ goes to zero, we found that
\begin{eqnarray}
\lim_{\rho \rightarrow 0} \left\{ h_G(\epsilon; \sigma_u^2) - h(\epsilon) \right\} \eqqq \frac{1}{2} \log_2 \frac{2 e \pi^3}{3} - \log_2 2 \pi  = \frac{1}{2}\log_2 \frac{e \pi}{6}, \\
\lim_{\rho \rightarrow 0} \left\{ h_V(\epsilon; \bar{R}_L) - h(\epsilon) \right\} \eqqq 0.
\end{eqnarray}
In other words, the von-Mises upperbound is asymptotically tight for sufficiently fast-varying channels, while the Gaussian upperbound has the constant gap. The asymptotical tightness of the derived upperbounds for sufficiently slow-varying channels, however, is hard to show since both of derived ones and the exact continuous entropy approach to negative infinity as $\rho$ goes to one. Let us denote $H(X)$ be the discrete version of $h(X)$. Then we can say that
\begin{eqnarray}
\lim_{\rho \rightarrow 1} H(\epsilon) = 0
\label{eq:H(eps) for rho 1}
\end{eqnarray}
since $\Pr_{\epsilon}(X) = \delta(X)$ holds in the case of (\ref{eq:asymp_exact1}). Also $\Pr_{\epsilon}(X) = \delta(X)$ holds both for the Gaussian PDF with zero variance (\ref{eq:asymp_Gauss1}) and the von-Mises PDF with unit MRL (\ref{eq:asymp_von1}). Thus, we can obtain
\begin{eqnarray}
\lim_{\rho \rightarrow 1} H_G(\epsilon; \sigma_u^2) = 0, \\
\lim_{\rho \rightarrow 1} H_V(\epsilon; \bar{R}_L) = 0,
\end{eqnarray}
and the asymptotic tightness of the discrete entropy of the derived upperbounds are shown as
\begin{eqnarray}
\lim_{\rho \rightarrow 1} \left\{ H_G(\epsilon; \sigma_u^2) - H(\epsilon) \right\} \eqqq 0,\\
\lim_{\rho \rightarrow 1} \left\{ H_V(\epsilon; \bar{R}_L) - H(\epsilon) \right\} \eqqq 0.
\end{eqnarray}

\section{Analysis of Feedback Overhead}
\subsection{Entropy and Feedback Overhead}
For MIMO beamforming under PAPC, a transmitter requires the phase steering vector $\bm{\theta}[n]$ to make the beamforming vector $\vb[n]$ in (\ref{def:v[n]}). Since the entropy determines the required amount of bits to convey the corresponding information, feedback overhead for MIMO beamforming can be represented by the entropy of the phase steering vector. Considering temporal correlation of fading channels, the phase tracking vector $\bm{\epsilon}[n]$ is sufficient to form the beamforming vector $\vb[n]$ based on the definition in (\ref{def:epsilon[n]}) assuming that the transmitter has prior information of $\bm{\theta}[n-1]$. Thus the entropy of the phase tracking vector becomes the required amount of feedback overhead when temporal correlation is taken into account to the encoding of feedback messages. The entropy of the phase steering vector is written as
\begin{eqnarray}
h\!\!\left(\bm{\theta}[n]\right) = (N_T - 1) \log_2 2\pi
\end{eqnarray}
due to {\em Theorem \ref{Theorem1}}.
On the other hand, the entropy of the phase tracking vector is upperbounded as
\begin{eqnarray}
h\!\!\left(\bm{\epsilon}[n]\right) \leq (N_T - 1) h(\epsilon)
\label{eq:ent_vector to scalar}
\end{eqnarray}
since the elements in the phase tracking vector may statistically correlated each other. Note that we have presented the upperbounds on $h(\epsilon)$ based on the theory of the maximum entropy distributions in Section IV.

In practical systems, feedback messages should be quantized to be delivered through limited feedback links. It is known \cite{Cover05} that $h(X) \rightarrow H(X) + \log_2 \Delta$ as $\Delta \rightarrow 0$ where the gap between two adjacent quantization levels is denoted by $\Delta$. Let $H\!\!\left(\bm{\theta}^{(L)}[n]\right)$ and $H\!\!\left(\bm{\epsilon}^{(L)}[n]\right)$ respectively denote the discrete entropies of the phase steering vector and the phase tracking vector where the elements of each vector are uniformly quantized into $L$-levels. Since the phase steering values are i.i.d. uniform RV's, the discrete entropy of the phase steering vector is written as
\begin{eqnarray}
H\!\!\left(\bm{\theta}^{(L)}[n]\right) = (N_T - 1) \log_2 L.
\end{eqnarray}
The discrete entropy of the phase tracking vector is approximately upperbounded as
\begin{eqnarray}
H\!\!\left(\bm{\epsilon}^{(L)}[n]\right) \lessapprox  (N_T - 1) \left(h(\epsilon) - \log_2 \frac{2\pi}{L}\right)
\end{eqnarray}
where $\lessapprox$ becomes $\leq$ for sufficiently large $L$.

\subsection{Mutual Information and Amount of Reduction in Feedback Overhead}
From (\ref{eq:h(eps[n])}), we know that $I\!\!\left( \bm{\theta}[n]; \bm{\theta}[n-1]\right) = h\!\!\left(\bm{\theta}[n]\right) - h\!\!\left(\bm{\epsilon}[n]\right)$ which represents the amount of reduction in feedback overhead for MIMO beamforming obtained from the temporal correlation property of fading channels. Let $I\!\!\left( \bm{\theta}^{(L)}[n]; \bm{\theta}^{(L)}[n-1]\right)$ also be the element-wise $L$-level uniform quantization version of $I\!\!\left( \bm{\theta}[n]; \bm{\theta}[n-1]\right)$, then from \cite{Cover05} we have $I\!\!\left( \bm{\theta}^{(L)}[n]; \bm{\theta}^{(L)}[n-1]\right) \approx I\!\!\left( \bm{\theta}[n]; \bm{\theta}[n-1]\right)$ where the approximation becomes accurate for sufficiently large $L$.

\begin{itemize}[\IEEEsetlabelwidth{Z}]
\item {\em Lowerbounds on MI with the Derived Results:} The MI between $\bm{\theta}[n]$ and $\bm{\theta}[n-1]$ is lowerbounded as
\begin{eqnarray}
I\!\!\left( \bm{\theta}[n]; \bm{\theta}[n-1]\right)
    \!\!\!\!&\underset{(a)}{\geq}&\!\!\!\!\ \left(N_T - 1\right) \left( \log_2 2\pi - h(\epsilon) \right)\nn\\
    \!\!\!\!&\underset{(b)}{\geq}&\!\!\!\!\ \left(N_T - 1\right) \left( \log_2 2\pi - h_u(\epsilon) \right)
\end{eqnarray}
where the inequality in (a) comes from the inequality in (\ref{eq:ent_vector to scalar}); $h_u(\epsilon)$ in (b) represents one of the derived upperbounds on the continuous entropy of the phase tracking value in the previous section, namely $h_G(\epsilon; \sigma_{\epsilon}^2)$, $h_G(\epsilon; \sigma_{u}^2)$, $h_V(\epsilon; \bar{R}_{\epsilon})$, and $h_V(\epsilon; \bar{R}_{L})$.

\item {\em Asymptotic Behavior of MI:} For sufficiently slow-varying channels, the MI between $\bm{\theta}[n]$ and $\bm{\theta}[n-1]$ becomes
\begin{eqnarray}
\lim_{\rho \rightarrow 1} I\!\!\left( \bm{\theta}[n]; \bm{\theta}[n-1]\right)
    \!\!\!\!&\underset{(a)}{\approx}&\!\!\!\!\
        \lim_{\rho \rightarrow 1} I\!\!\left( \bm{\theta}^{(L)}[n]; \bm{\theta}^{(L)}[n-1]\right) \nn\\
    \!\!\!\!&\underset{(b)}{=}&\!\!\!\!\
        H\!\!\left(\bm{\theta}^{(L)}[n]\right) - \lim_{\rho \rightarrow 1} H\!\!\left(\bm{\epsilon}^{(L)}[n]\right) \nn\\
    \!\!\!\!&\underset{(b)}{=}&\!\!\!\!\
        (N_T - 1) \log_2 L
\end{eqnarray}
where the approximation in (a) becomes accurate for large $L$; (b) follows from the fact that $H\!\!\left(\bm{\theta}^{(L)}[n]\right) = (N_T - 1) \log_2 L$ which is not a function of temporal correlation parameter $\rho$; (c) follows from the fact that $H\!\!\left(\bm{\epsilon}^{(L)}[n]\right)$ goes to zero as $\rho \rightarrow 1$ as shown in (\ref{eq:H(eps) for rho 1}). On the otherhand, for sufficiently fast-varying channels, the MI becomes
\begin{eqnarray}
\lim_{\rho \rightarrow 0} I\!\!\left( \bm{\theta}[n]; \bm{\theta}[n-1]\right)
    \!\!\!\!&\underset{(a)}{=}&\!\!\!\!\
        h\!\!\left(\bm{\theta}[n]\right) - \lim_{\rho \rightarrow 0} h\!\!\left(\bm{\epsilon}[n]\right) \nn\\
    \!\!\!\!&\underset{(b)}{=}&\!\!\!\!\
        h\!\!\left(\bm{\theta}[n]\right) - \lim_{\rho \rightarrow 0} h\!\!\left(\bm{\theta}[n] | \bm{\theta}[n-1]\right) \nn\\
    \!\!\!\!&\underset{(c)}{=}&\!\!\!\!\
        h\!\!\left(\bm{\theta}[n]\right) - h\!\!\left(\bm{\theta}[n]\right) = 0
\end{eqnarray}
where (a) comes from the fact that $h\!\!\left(\bm{\theta}[n]\right)$ is not a function of $\rho$, (b) comes from the relation $h\!\!\left(\bm{\epsilon}[n]\right) = h\!\!\left(\bm{\theta}[n] | \bm{\theta}[n-1]\right)$ in (\ref{eq:h(eps[n])}); (c) from the distribution equality in (\ref{eq:Distribution_equality}), we know that as $\rho$ goes to zero, $\gamma$ goes to infinity. Thus $\theta_l[n]$ becomes independent to $\theta_l[n-1]$ when we consider the last equality in (\ref{eq:theta_l[n]3}).
\end{itemize}

\section{Discussion and Numerical Results}

\subsection{Comparison of Entropies}

\begin{figure*}[!t]
\centerline{\subfloat[]{\includegraphics[width=3.5in]{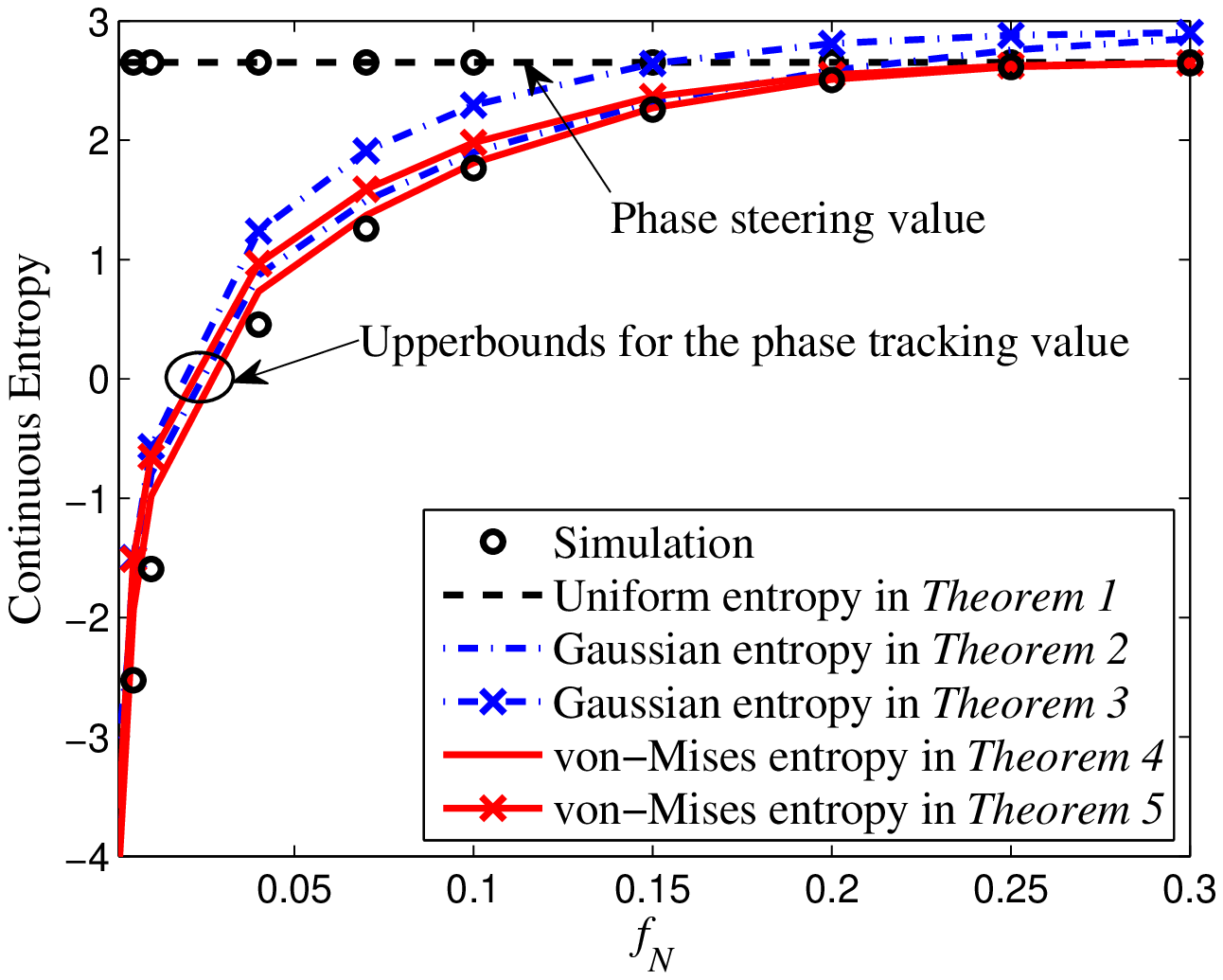}
\label{fig:C_entropy}}
\hfil
\subfloat[]{\includegraphics[width=3.5in]{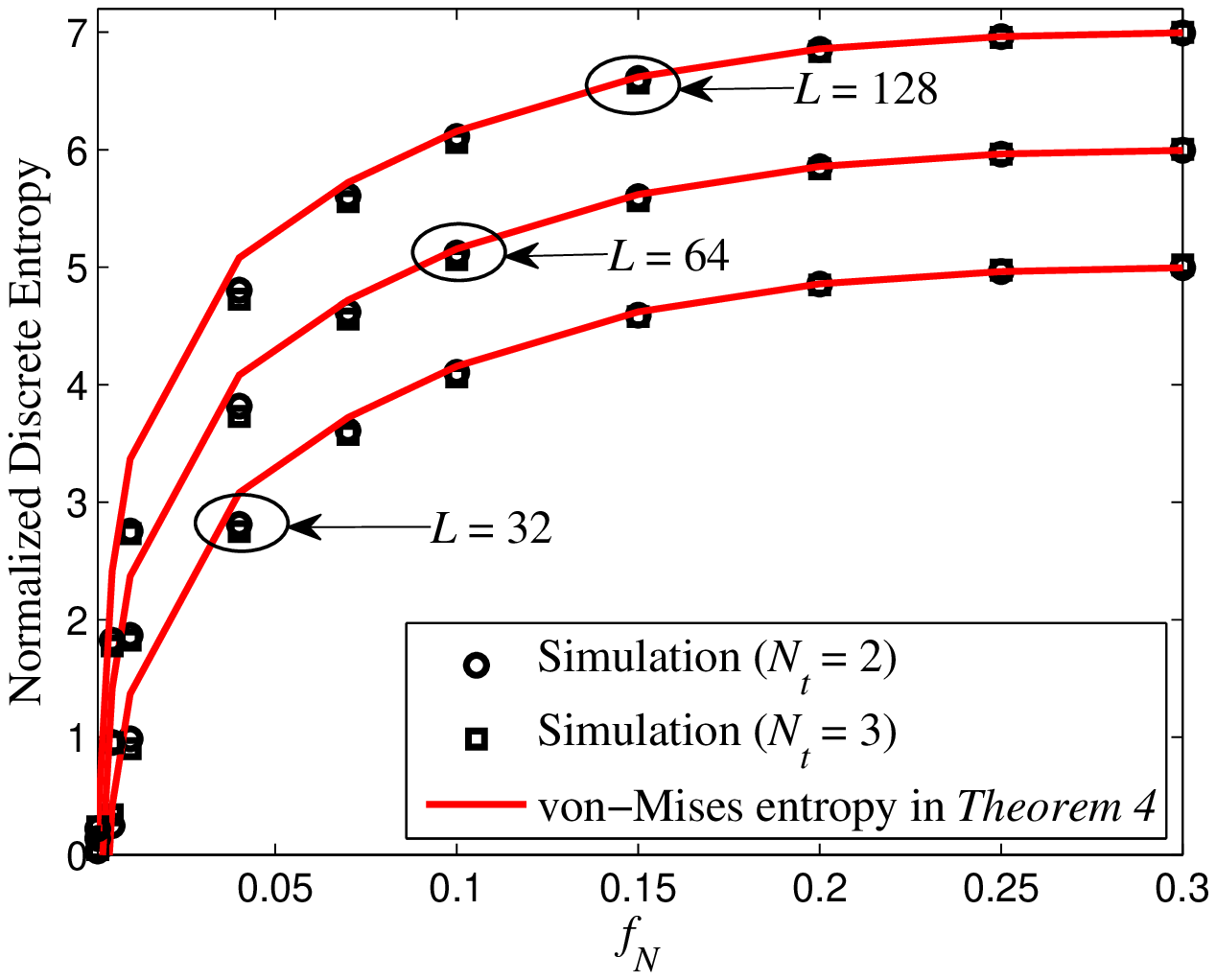}
\label{fig:D_entropy}}}
\caption{Comparison of feedback overhead for the phase steering and tracking values in terms of (a) continuous entropy and (b) discrete entropy.}
\label{fig:entropies}
\end{figure*}

In Fig.~\ref{fig:entropies}, continuous and discrete entropies of the phase steering and tracking values are compared. Simulation results for the discrete entropies are obtained via Monte-Carlo simulation. Both of the phase steering and tracking values are uniformly quantized to $L=128$ levels. Corresponding simulation results for the continuous entropies are transformed by using the relation $h(X) \approx H(X^{(L)}) + \log_2 \frac{2\pi}{L}$. The results show that the continuous and discrete entropies of the phase tracking value are monotonic increasing functions of the NDF while the ones of the phase steering value are constant.

In Fig.~\ref{fig:C_entropy}, the derived curves of continuous entropies for the phase steering and tracking values are compared to the simulation results. In the case of phase steering values, the simulation result is exactly matched to the continuous entropy $\log_2 2\pi$ in {\em Theorem 1}. In the case of phase tracking value, the results of analytical upperbound in {\em Theorems 2, 3, 4,} and 5 are jointly plotted. The von-Mises upperbounds in {\em Theorem 4} and {\em Theorem 5} are more accurate than the ones for the Gaussian upperbound in {\em Theorem 2} and {\em Theorem 3}. Note that the upperbounds in {\em Theorem 3} and {\em Theorem 5} are closed-form expressions which are relatively looser than the corresponding numerical upperbounds in {\em Theorem 2} and {\em Theorem 4}. For high NDF regions, the gap between the von-Mises upperbounds and the simulation result approaches to zero while the Gaussian upperbounds and the simulation result has constant gap as mentioned in Section III-C.

In Fig.~\ref{fig:D_entropy}, discrete entropies of the phase tracking vector normalized by the number of tracking values in the vector, i.e., $\frac{H\!\!\left(\bm{\epsilon}^{(L)}[n]\right)}{N_T-1}$, are depicted for $L = 32, 64,$ and $128$. As $L$ (the number of element-wise uniform quantization levels) increases, the required amount of feedback overhead also increases since information loss caused by quantization decreases. The von-Mises closed-form upperbound in {\em Theorem 4} is plotted which is the most tight one with the simulation result among the four upperbounds presented above. The correlation between elements in the phase tracking vector can be verified in this figure by inspecting the gap between simulation results for $N_T = 2$ and $N_T = 3$. The gap represents normalized MI between elements in the phase tracking vector , i.e., $\frac{I\left( {\epsilon_1}[n]; {\epsilon_2}[n]\right)}{2}$, which is not zero though it seems negligibly small. Note that the upperbound is tight to the simulation results both for sufficiently slow-varying and fast-varying channels.

\begin{figure*}[!t]
\centerline{\subfloat[]{\includegraphics[width=3.5in]{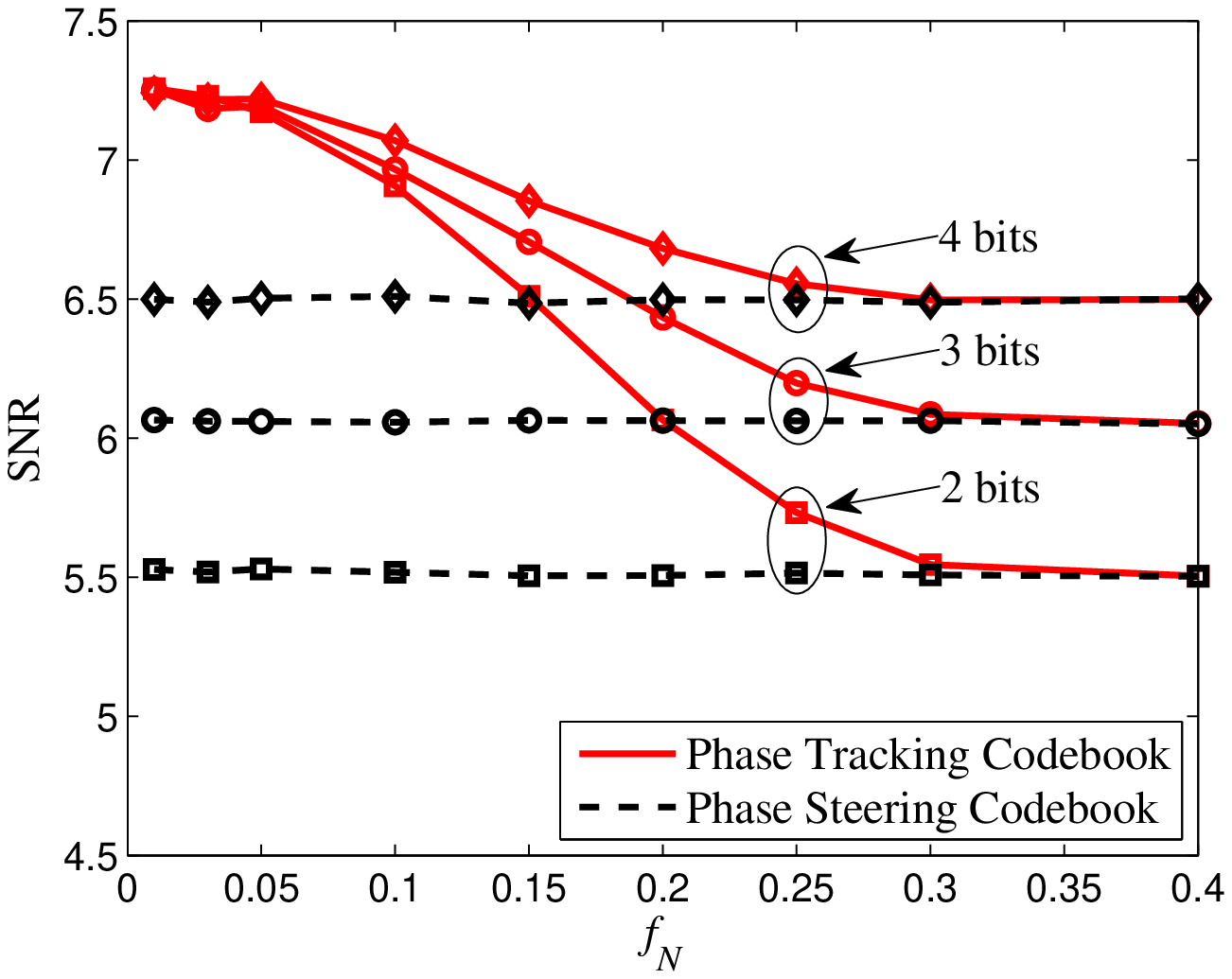}
\label{fig:CB_perf}}
\hfil
\subfloat[]{\includegraphics[width=3.5in]{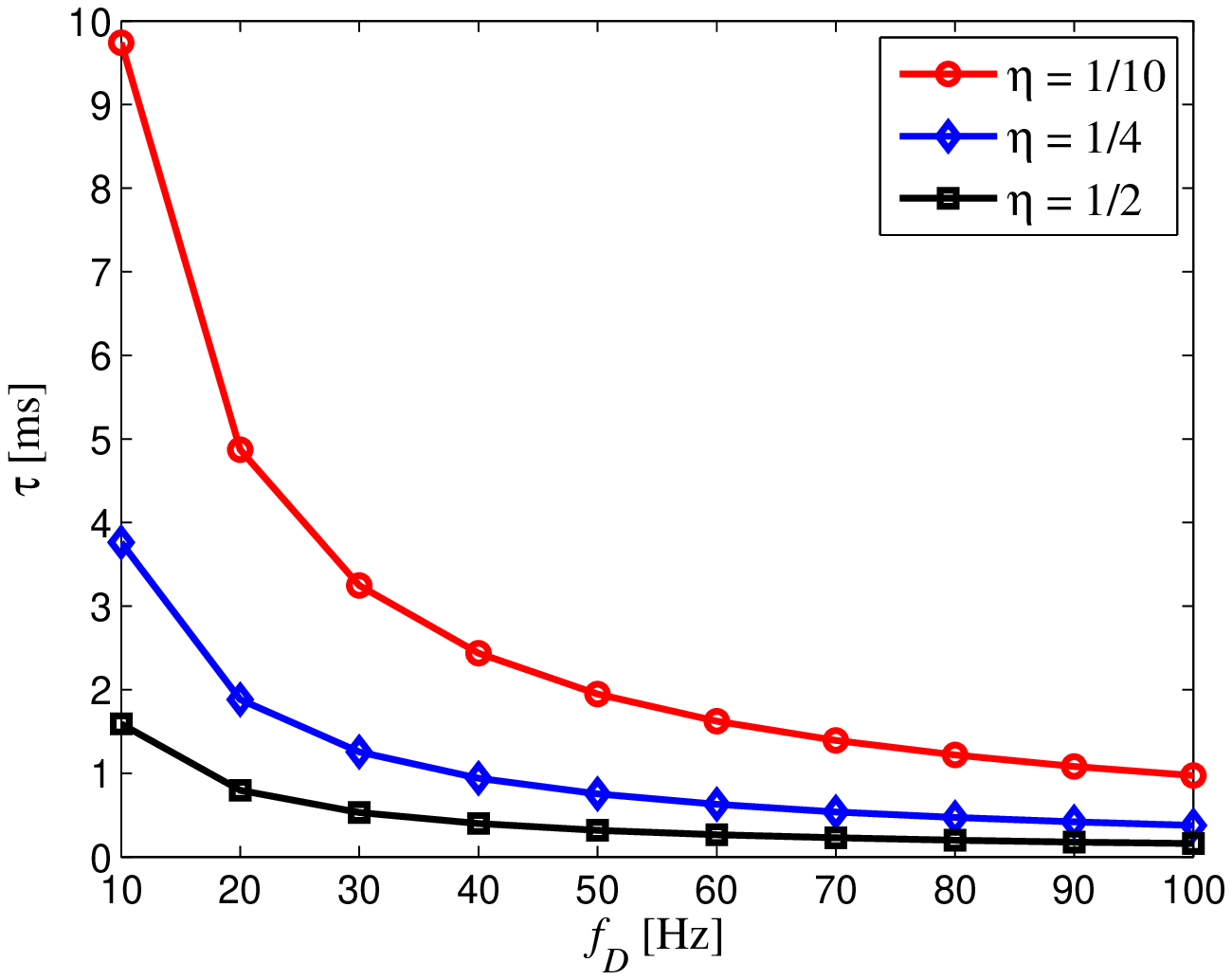}
\label{fig:FB_duration}}}
\caption{ Applications (a) Phase tracking codebook outperforms phase steering codebook with smaller number of feedback overheads (b) Determination of feedback duration in the sense that MI between adjacent feedback slots should larger than certain threshold $\eta$}
\label{fig:applications}
\end{figure*}

\subsection{Applications}

\subsubsection{Phase Tracking Codebook}
We construct phase steering and tracking codebooks to evaluate the average SNR performance w.r.t. the amount of feedback overhead. The phase steering codebook is simply generated by element-wise $L$-level uniform quantization to the phase steering vector. Hence $(N_T -1) \log_2 L$ becomes the feedback overhead of the phase steering codebook. The phase tracking codebook is generated via generalized Lloyd type vector quantization (VQ) technique \cite{Park10}. Letting the size of the codebook be $T$, the feedback overhead of the phase tracking codebook becomes $\log_2 T$. The simulated average SNR results are depicted in Fig.~\ref{fig:CB_perf}. Unlike the phase steering codebook, the performance of the tracking codebook is a function of $f_N$ which improves more as the NDF become smaller. The entropy of the phase tracking vector decreases for small $f_N$, and thus the phase tracking codebook can convey more information of feedback messages as NDF become smaller though the amount of feedback overhead remains constant. As shown in Fig.~\ref{fig:CB_perf}, the phase tracking codebook can outperform the phase steering codebook even when the amount of feedback information bit is significantly smaller when the degree of channel variation is sufficiently small.

\subsubsection{Influence to the Feedback Duration}

Let $I\!\!\left( \bm{\theta}[n]; \bm{\theta}[n-1]\right) = \eta h\!\!\left(\bm{\theta}[n]\right)$ where $\eta \in [0,~1]$ is the threshold of remainder information of phase steering values at the transmitter after the duration of feedback slots $\tau$. This also can be represented as
\begin{eqnarray}
h\!\!\left(\bm{\epsilon}[n]\right) \eqqq (1 - \eta) h\!\!\left(\bm{\theta}[n]\right)\nn\\
    \eqqq (1 - \eta) (N_T - 1 ) \log_2 2 \pi.
\label{eq:FB_duration}
\end{eqnarray}
Now to determine the appropriate feedback duration, firstly $\hat{\rho}$ should be found which satisfies (\ref{eq:FB_duration}). Then by using $\hat{\rho}$ and the relation in (\ref{def:rho}), we can obtain $\hat{f}_N$ and corresponding $\hat{\tau} = \hat{f}_N / f_D$. Note that $\hat{\tau}$ represents the appropriate duration of feedback slots which guarantees the quality of the outdated feedback messages. In Fig.~\ref{fig:FB_duration}, the appropriate feedback duration $\hat{\tau}$ is depicted for $\eta = 0.1, 0.25,$ and $0.5$ w.r.t. the maximum Doppler frequency $f_D$, where instead of using $h\!\!\left(\bm{\epsilon}[n]\right)$ in (\ref{eq:FB_duration}) the closed-form von-Mises upperbound in {\em Theorem 5} is used. As $\tau$ increases the appropriate duration $\hat{\tau}$ decreases since the required amount of the remainder information in the outdated feedback messages also increases. The appropriate duration $\hat{\tau}$ also decreases as the maximum Doppler frequency $f_D$ increases since the temporal correlation parameter $\rho$ and corresponding $h\!\!\left(\bm{\epsilon}[n]\right)$ also become small.

\section{Conclusion}
The feedback overhead for MIMO beamforming is analyzed considering temporal correlation property of the wireless fading channels. Under PAPC, two types of feedback messages, the phase steering and tracking values, are investigated and the corresponding entropies are derived in the functions of temporal correlation parameter. For the phase steering value, the i.i.d. property is proved in {\em Theorem 1}. For the phase tracking values, four upperbound entropies are derived based on the Gaussian PDF and von-Mises PDF in {\em Theorem 2 -- Theorems 5} where two of them in {\em Theorems 3, 5} are closed-form expressions. The derived results can represent the amount of required bits for each type of feedback message via rate-limited feedback links and also show the amount of reduction in feedback overhead obtained from taking temporal correlation property into account for encoding of feedback messages. For application perspective, the phase tracking codebook generated by VQ algorithm is proposed which can outperform the phase steering codebook with smaller number of feedback bits. Also the derived results can be used to determine the appropriate duration of feedback report with the function of the threshold of remainder information at transmitter and the maximum Doppler frequency.

\appendix

\subsection{Relation between the Continuous Entropy of the von-Mises PDF and the Corresponding MRL}

The continuous entropy of von-Mises PDF is inversely proportional to the corresponding MRL.
It is directly proved from following two {\em Remarks}.
\begin{remark}
The MRL of von-Mises PDF $\bar{R} = \frac{I_1(\kappa)}{I_0(\kappa)}$ is a monotonic increasing function w.r.t. the concentration parameter $\kappa$.
\label{Lemma:MRL_and_CP}
\end{remark}
\begin{IEEEproof}
Let the ratio of two modifided Bessel functions of the first kind be denoted by
\begin{eqnarray}
r_{\nu}(x) = \frac{I_{\nu+1}(x)}{I_{\nu}(x)}
\label{def:ratio_of_Bessels}
\end{eqnarray}
for $\nu \geq 0$. Then the ratio can be rewritten as
\begin{eqnarray}
r_{\nu}(x) = \frac{x}{ \nu + 1 + \left(R_{\nu+1}x^2 + (\nu + 1)^2\right)^{\frac{1}{2} }}.
\end{eqnarray}
where $R_{\nu+1} = r_{\nu+1}/r_{\nu}$ \cite{Amos74}. Note that $0 \leq R_{\nu+1} \leq 1$ holds since $0 \leq r_{\nu+1}(x) \leq r_{\nu}(x) < 1$ is satisfied \cite{Amos74}. Taking derivative to the ratio, we derive
\begin{eqnarray}
\frac{d}{dx}r_{\nu}(x) = \frac{1}{ 1 + \nu + \frac{R_{\nu+1}x^2}{1+\nu} + \sqrt{(1+\nu)^2 + R_{\nu+1}x^2}   }.
\end{eqnarray}
Since $\nu \geq 0$ and $0 \leq R_{\nu+1} \leq 1$, the derivative $\frac{d}{dx}r_{\nu}(x)$ is always positive regardless of the value of $\nu$.
\end{IEEEproof}
\begin{remark}
The continuous entropy of von-Mises PDF is a monotonic decreasing function w.r.t. the concentration parameter $\kappa$.
\label{Lemma:Ent and CP}
\end{remark}
\begin{IEEEproof}
By using (\ref{def:ratio_of_Bessels}), the continuous entropy of von-Mises PDF in (\ref{def:von-Mises entropy}) is rewritten as
\begin{eqnarray}
h(V) = -{\kappa} r_{0}(\kappa) \log_2 e + \log_2 (2\pi I_0({\kappa})).
\end{eqnarray}
Differentiating w.r.t. $\kappa$, we obtain
\begin{eqnarray}
\frac{d}{d\kappa}h(V) = -r_{0}(\kappa)\log_2 e - {\kappa} r_{0}'(\kappa)\log_2 e + \frac{I_0'({\kappa})}{I_0({\kappa})}\log_2 e.
\end{eqnarray}
By using the facts that $I_{\nu}'(x)/I_{\nu}(x) = r_{\nu}(x) + \nu/x$ and $0 < r_{\nu}'(x)$ in \cite{Amos74}, we derive
\begin{eqnarray}
\frac{d}{d\kappa}h(V) \eqqq \left(-r_{0}(\kappa) - {\kappa} r_{0}'(\kappa) + r_{0}(\kappa)\right)\log_2 e\nn\\
    \eqqq - {\kappa} r_{0}'(\kappa)\log_2 e \leq 0.\nn
\end{eqnarray}
\end{IEEEproof}

\subsection{Proof of Lemma \ref{Lemma:LB_MRL}}
For the range of $0 \leq \gamma \leq 1$, the conditional expectation $\bar{C}(\gamma)$ is lowerbounded as
\begin{eqnarray}
\bar{C}_{\epsilon}(\gamma)
    \eqqq \frac{1}{2\pi} \int_{-\pi}^{\pi} \frac{1 + \gamma \cos\psi}{\sqrt{1 + \gamma^2 + 2\gamma \cos \psi}} d\psi \nn\\
       \!\!\!\!&\underset{(a)}{\geq}&\!\!\!\!
   \frac{1}{2\pi} \int_{-\pi}^{\pi} \frac{(1 + \gamma\cos\psi)^2}{{1 + \gamma^2 + 2\gamma\cos\psi}} d\psi \nn\\
        \eqqq 1 - \frac{1}{2\pi} \int_{-\pi}^{\pi} \frac{\gamma^2\sin^2\psi}{{1 + \gamma^2 + 2\gamma\cos\psi}} d\psi \nn\\
        \eqqq 1 - \frac{\gamma^2}{2} \nn\\
        \!\!\!\!&{\triangleq}&\!\!\!\! \bar{C}_{L_1}(\gamma)
\label{eq:LB1_C_gamma}
\end{eqnarray}
where $\frac{1 + \gamma \cos\psi}{\sqrt{1 + \gamma^2 + 2\gamma \cos \psi}} \leq 1$ holds for $0 \leq \gamma \leq 1$ due to $(1+\gamma \cos\psi)^2 \leq 1 + \gamma^2 + 2\gamma \cos\psi$, and (a) follows from the fact that $x^2 \leq x$ holds for $0 \leq x \leq 1$.

For the range of $\gamma > 1$,
\begin{eqnarray}
\bar{C}_{\epsilon}(\gamma)
    \eqqq \frac{1}{2\pi} \int_{-\pi}^{\pi} \frac{1 + \gamma \cos\psi}{\sqrt{1 + \gamma^2 + 2\gamma \cos \psi}} d\psi \nn\\
    \!\!\!\!&\underset{(a)}{=}&\!\!\!\!
        \frac{1}{\pi} \int_{0}^{\pi} \frac{1 + \gamma \cos\psi}{\sqrt{1 + \gamma^2 + 2\gamma \cos \psi}} d\psi \nn\\
    \!\!\!\!&\underset{(b)}{=}&\!\!\!\!
        \frac{1}{\pi} \int_{0}^{\psi_T} \frac{1 + \gamma \cos\psi}{\sqrt{1 + \gamma^2 + 2\gamma \cos \psi}} d\psi
           + \frac{1}{\pi} \int_{\psi_T}^{\pi} \frac{1 + \gamma \cos\psi}{\sqrt{1 + \gamma^2 + 2\gamma \cos \psi}} d\psi
\label{eq:LB1b}
\end{eqnarray}
where (a) follows from the fact that $\cos(x) = \cos(-x)$ and $\psi_T$ in (b) satisfies $\gamma\cos\psi_T = -1$. Since $1+\gamma\cos\psi$ has a positive value when $\psi \in [0,~\psi_T)$ and a negative value when $\psi \in [\psi_T, \pi]$, the $\bar{C}_{\epsilon}(\gamma)$ is lowerbounded as
\begin{eqnarray}
\bar{C}_{\epsilon}(\gamma)
    \!\!\!\!&{\geq}&\!\!\!\!
        \frac{1}{\pi} \int_{0}^{\psi_T} \frac{1 + \gamma \cos\psi}{\sqrt{1 + \gamma^2 + 2\gamma}} d\psi
           + \frac{1}{\pi} \int_{\psi_T}^{\pi} \frac{1 + \gamma \cos\psi}{\sqrt{1 + \gamma^2 - 2\gamma}} d\psi \nn \\
    \eqqq \frac{\psi_T + \gamma \sin\psi_T}{\pi(\gamma + 1)} + \frac{\pi - (\psi_T + \gamma \sin\psi_T)}{\pi(\gamma - 1)} \nn \\
    \eqqq \frac{1}{(\gamma - 1)}  - \frac{2(\psi_T + \gamma \sin\psi_T )}{\pi(\gamma + 1)(\gamma - 1)}.
\label{eq:LB2b}
\end{eqnarray}
Since $\gamma\cos\psi_T = -1$, we have $\cos^2\psi_T = 1/\gamma^2$ and $\sin^2\psi_T = 1 - 1/\gamma^2$ to satisfy $\sin^2\psi_T + \cos^2\psi_T = 1$. Thus $\gamma^2\sin^2\psi_T = \gamma^2 - 1$ is obtained. Further since positiveness of $\gamma^2\sin^2\psi_T$ is guaranteed for $\psi_T \in [0,~\pi]$, the equality $\gamma\sin\psi_T = \sqrt{\gamma^2 - 1}$ holds. Hence (\ref{eq:LB2b}) can be written as
\begin{eqnarray}
\bar{C}_{\epsilon}(\gamma)
    \!\!\!\!&{\geq}&\!\!\!\!
        \frac{1}{(\gamma - 1)}  - \frac{2\psi_T}{\pi(\gamma + 1)(\gamma - 1)} - \frac{2}{\pi\sqrt{\gamma^2 - 1}}.
\label{eq:LB2c}
\end{eqnarray}
Also since $\gamma\cos\psi_T = -1$, we have $\cos(\pi - \psi_T) = 1/\gamma$. By using the trigonometry identities, $\tan(\pi - \psi_T) = \gamma^2 - 1$ is obtained, and thus $\psi_T = \pi - \tan^{-1}(\gamma^2 - 1)$ can be derived. Hence (\ref{eq:LB2c}) can be rewritten as
\begin{eqnarray}
\bar{C}_{\epsilon}(\gamma)
    \!\!\!\!&{\geq}&\!\!\!\!
        \frac{1}{(\gamma - 1)}  - \frac{2(\pi - \tan^{-1}(\gamma^2 - 1))}{\pi(\gamma + 1)(\gamma - 1)} - \frac{2}{\pi\sqrt{\gamma^2 - 1}}\nn\\
    \eqqq \frac{1}{(\gamma + 1)} + \frac{2\tan^{-1}\sqrt{\gamma^2 - 1}}{\pi(\gamma + 1)(\gamma - 1)}
        + \frac{-2}{\pi\sqrt{\gamma^2 - 1}} \nn\\
    \!\!\!\!&{\triangleq}&\!\!\!\! \bar{C}_{L_2}(\gamma)
\label{eq:LB2_C_gamma}
\end{eqnarray}

\subsection{Proof for Non-negativeness of $c_i(k)$ in (\ref{eq:Exp_LB_C})}

Firstly, $c_1(k) \geq 0$ directly follows from the fact that $\bar{C}_{L_1}(\gamma) \geq 0$ for $0 \leq \gamma \leq 1$. Secondly, $c_2(k) \geq 0$ is proved by showing that $\bar{C}_{L_2}(x) \geq 0$ due to the definition $c_2(k) = \int_1^{\infty} \bar{C}_{L_2}(x) f_{\gamma}(x)dx$ in (\ref{eq:Exp_LB_C}). Note that
\begin{eqnarray}
\bar{C}_{L_2}(\gamma)
    \eqqq \frac{1}{(\gamma + 1)} + \frac{2\tan^{-1}\sqrt{\gamma^2 - 1}}{\pi(\gamma + 1)(\gamma - 1)} + \frac{-2}{\pi\sqrt{\gamma^2 - 1}} \nn\\
    \!\!\!\!&\underset{(a)}{\geq}&\!\!\!\!
        \frac{1}{(\gamma + 1)} + \frac{6 \sqrt{\gamma^2 - 1}}{\pi(\gamma + 1)(\gamma - 1)(1+2\gamma)} + \frac{-2}{\pi\sqrt{\gamma^2 - 1}} \nn\\
    \eqqq
        \frac{\pi\sqrt{\gamma^2-1}(\gamma-1)(1+2\gamma) - 4(\gamma + 1)(\gamma-1)^2}{\pi \sqrt{\gamma^2-1}(\gamma-1) (\gamma+1) (1+2\gamma)} \triangleq \bar{C}_{a}(\gamma)
\end{eqnarray}
where (a) follows from the inequality $\tan^{-1}(x) > \frac{3x}{1+2\sqrt{1+x^2}}$ in \cite{Thorp66}. Since the numerator of $\bar{C}_{a}(\gamma)$ is greater or equal to zero for $\gamma > 1$, we need to show that the denominator of $\bar{C}_{a}(\gamma)$ is also greater or equal to zero for the same range of $\gamma$. The denominator of $\bar{C}_{a}(\gamma)$ is lowerbounded by ${\pi(\gamma-1)^2(1+2\gamma) - 4(\gamma + 1)(\gamma-1)^2} \triangleq \bar{C}_{b}(x)$ since the inequality $\sqrt{\gamma^2 - 1} \geq \gamma -1$ holds for $\gamma > 1$. Differentiating $\bar{C}_{b}(\gamma)$ w.r.t. $\gamma$, we obtain $\frac{d}{d\gamma}\bar{C}_{b}(\gamma) = 6(\pi-1)\gamma^2 - 2(3\pi - 4)\gamma +4$ which has two roots at $\gamma = 1$ and $\gamma = \frac{2}{3(\pi-2)}$. Since $\bar{C}_{b}(x) = 2(\pi-2)\gamma^3 + \textrm{O}(\gamma^2)$ and $2(\pi-2) > 0$, the local minimum of $\bar{C}_{b}(\gamma)$ occurs at $\gamma = 1$ and $\bar{C}_{b}(\gamma) > \bar{C}_{b}(1) = 0$ for all $\gamma > 1$. Thus the positiveness of $\bar{C}_{a}(\gamma)$ is guaranteed since the denominator of $\bar{C}_{a}(\gamma)$ is lowerbounded by $\bar{C}_{b}(\gamma) > 0$.


\begin{thebibliography}{10}

\bibitem{Goldsmith05} A. Goldsmith, {\em Wireless Communications}. Cambridge, U.K.: Cambridge Univ. Press, 2005.

\bibitem{Au-Yeung07} C. K. Au-Yeung and D. J. Love, ``{On the performance of random vector quantization limited feedback beamforming in a MISO system},'' {\em IEEE Trans. Wireless Commun.}, vol.~6, no.~2, pp.~458--462, Feb. 2007.
\bibitem{Roh06} J. C. Roh and B. D. Rao, ``{Design and analysis of MIMO spatial multiplexing systems with quantized feedback},'' {\em IEEE Trans. Signal Process.}, vol.~54, no.~8, pp.~2874--2886, Aug. 2006.
\bibitem{Love03G} D. J. Love and R. W. Heath, Jr., ``{Grassmannian BF for multiple-input multiple output systems},'' {\em IEEE Trans. Inform. Theory}, vol.~49, no.~10, pp.~2735--2747, Oct. 2003.
\bibitem{Love05} D. J. Love and R. W. Heath, Jr., ``{Limited feedback unitary precoding for spatial multiplexing systems},'' {\em IEEE Trans. Inform. Theory}, vol.~51, no.~8, pp.~2967--2976, Aug. 2005.

\bibitem{Yu07} W. Yu and T. Lan, ``{Transmitter optimization for the multi-antenna downlink with per-antenna power constraints},'' {\em IEEE Trans. Signal Process.}, vol.~55, no.~6, pp.~2646--2660, June 2007.

\bibitem{Love08} D.J. Love, R. W. Heath, Jr., V. K. N. Lau, D. Gesbert, B. D. Rao, and M. Andrews, ``{An overview of limited feedback in wireless communication systems},'' {\em IEEE J.
    Select. Areas Commun.}, vol.~26, no.~8, pp.~1341--1365, Oct. 2008.

\bibitem{Murthy07} C. R. Murthy and B. D. Rao, ``{Quantization methods for equal gain transmission with finite rate feedback},'' {\em IEEE Trans. Signal Process.}, vol.~55, no.~1, pp.~233--245, Jan. 2007.
\bibitem{Love03E} D. J. Love and R. W. Heath, Jr., ``{Equal gain transmission in multiple-input multiple-output wireless systems},'' {\em IEEE Trans. Wireless Commun.}, vol.~51, no.~7, pp.~1102--1120, July 2003.

\bibitem{Heath98} R. W. Heath, Jr. and A. Paulraj, ``{A simple scheme for transmit diversity using partial channel feedback},'' in {\em Proc. of IEEE Asilomar Conf. on Signals, Systems, and Comp.}, vol.~2, Nov. 1998, pp.~1073--1078.
\bibitem{Lee09} J. Lee, R. U. Naber, J. P. Choi, and H.-L. Lou, ``{Generalized co-phasing for multiple transmit and receive antennas},'' {\em IEEE Trans. Wireless Commun.}, vol.~8, no.~4, pp.~1649--1654, April 2009.
\bibitem{Zheng07} X. Zheng, Y. Xie, J. Li, and P. Stoica, ``{MIMO transmit beamforming under uniform elemental power constraint},'' {\em IEEE Trans. Signal Process.}, vol.~55, no.~11, pp.~5395--5406, Nov. 2007.
\bibitem{Hochwald00} B. M. Hochwald, T. Marzetta, T. Richardson, W. Sweldens, and R. Urbanke, ``{Systematic design of unitary space-time constellations},'' {\em IEEE Trans. Inform Theory}, vol.~46, no.~6, pp.~1962--1973, Sep. 2000.
\bibitem{Xia05} P. Xia, S. Zhou, and G. B. Giannakis, ``{Achieving the Welch bound with difference sets},'' {\em IEEE Trans. Inform Theory}, vol.~51, no.~5, pp.~1900--1907, May 2005.

\bibitem{Jakes74} W. C. Jakes, {\em Microwave mobile communications}, New York: Wiley, 1974.
\bibitem{Koorapaty05} H. Koorapaty, L. Krasny, and R. Ram\'{e}sh, ``{Delta modulation for channel feedback in transmit diversity systems},'' in {\em Proc. of IEEE Veh. Technol. Conf.}, vol.~1, pp.644--648, May 2005.
\bibitem{Chin08} W. H. Chin and C. Yuen, ``{Design of differential quantization for low bitrate channel state information feedback in MIMO-OFDM systems},'' in {\em Proc. of IEEE Veh. Technol. Conf.}, pp.~827--831, May 2008.
\bibitem{Cho05} M. Cho, W. Seo, Y. Kim, and D. Hong, ``{A joint feedback reduction scheme using delta modulation for dynamic channel allocation in OFDMA systems},'' in {\em Proc. of PIMRC}, vol.~4, , pp.~2747--2750, Sept. 2005.
\bibitem{Roh07} J. C. Rho and B. D. Rao, ``{Efficient feedback methods for MIMO channels based on parameterization},'' {\em IEEE Trans. Wireless Commun.}, vol.~6, no.~1, pp.~282--292, Jan. 2007.
\bibitem{Huang09} K. Huang, R. W. Heath, Jr., and J. G. Andrews, ``{Limited feedback beamforming over temporally-correlated channels},'' {\em IEEE Trans. Signal Process.}, vol.~57, no.~5, pp.~1--18, May 2009.
\bibitem{Mondal06} B. Mondal and R. W. Heath, Jr., ``{Channel adaptive quantization for limited feedback MIMO beamforming systems},'' {\em IEEE Trans. Signal Process.}, vol.~54, no.~12, pp.~4717--4729, Dec. 2006.
\bibitem{Samanta05} R. Samanta and R. W. Heath, Jr., ``{Codebook adaptation for quantized MIMO beamforming systems},'' in {\em Proc. of IEEE Asilomar Conf. on Signals, Systems, and Comp.}, pp.~376--380, Oct./Nov. 2005.

\bibitem{Sorrentino08} S. Sorrentino and L. Morreti, ``{Efficient precoder quantization for time-varying MIMO wireless channels},'' in {\em Proc. of IEEE Veh. Technol. Conf.}, pp.~787--791, May 2008.
\bibitem{Banister03a} B. C. Banister and J. R. Zeidler, ``{A simple gradient sign algorithm for transmit antenna weight adaptation with feedback},'' {\em IEEE Trans. Signal Process.}, vol.~51, no.~5, pp.~1156--1171, May 2003.
\bibitem{Banister03b} B. C. Banister and J. R. Zeidler, ``{Feedback assisted transmission subspace tracking for MIMO systems},'' {\em IEEE J. Select. Areas Commun.}, vol.~21, no.~3, pp.~452--463, April 2003.
\bibitem{Yang07} J. Yang and D. B. Williams, ``{Transmission subspace tracking for MIMO systems with low-rate feedback},'' {\em IEEE Trans. Commun.}, vol.~55, no.~8, pp.~1629--1639, Aug. 2007.
\bibitem{Murga09a} D. Sacrist\'{a}n-Murga and A. Pascual-Iserte, ``{Differential feedback of MIMO correlation matrices based on geodesic curves},'' in {\em Proc. of IEEE International Conference on Accoustics, Speech and Signal Processing (ICASSP 2009)}, Taipei, Taiwan, April 2009.
\bibitem{Murga09b} D. Sacrist\'{a}n-Murga, F. Kaltenberger, A. Pascual-Iserte, and A. I. P\'{e}rez-Neira, ``{Differential feedback in MIMO communications: performance with delay and real channel measurements},'' in {\em Workshop on Smart Antennas (WSA 2009)}, Berlin, Germany, Feb. 2009.

\bibitem{Cover05} T. M. Cover and J. A. Thomas, {\em Elements of information theory}, 2nd ed. New Jersey: Wiley 2005
\bibitem{Mardia72} K. V. Mardia, {\em Statistics of directional data}, New York: Academic, 1972.
\bibitem{Abramowitz64} M. Abramowitz and I. A. Stegun {\em Handbook of mathematical functions}, Washington, DC: Nat. Bur. Stand., 1964.
\bibitem{Amos74} D. E. Amos, ``{Computation of modified Bessel functions and theis ratios},'' {\em Math. Comp.}, vol.~28, no.~125, pp.~239--251, Jan. 1974.
\bibitem{Thorp66} E. O. Thorp, M. Fried, R. E. Shafer, et al., ``{Problems and solutions: elementary problems: E1865--E1874},'' {\em The American Mathematical Monthly}, vol.~73, no.~3, pp.~309--310,~1966.
\bibitem{Park10} J. Park, J. Kim, H. Yoo, and W. Sung, ``{Construction on phase tracking codebook based on Lloyd-Max vector quantization},'' submitted to {\em IEEE Global Telecommun. Conf. (GLOBECOM)} 2010.



\end{thebibliography}
\end{document}